\documentclass[journal]{IEEEtran}
\usepackage{graphicx}
\graphicspath{{tupian/}}
\usepackage{aligned-overset}
\usepackage{amstext}

\usepackage{mathrsfs}
\usepackage{amsfonts}
\usepackage{caption}
\usepackage{algorithm}
\usepackage{algpseudocode}
\usepackage{array}
\usepackage{booktabs}
\usepackage{amsmath}
\usepackage{subfigure}   
\usepackage{url}
\usepackage{amssymb}
\usepackage{framed}
\usepackage{color}
\usepackage[toc]{multitoc}
\usepackage{blindtext}
\usepackage{diagbox}
\usepackage{comment}
\usepackage{multirow}
\usepackage{wasysym}
\usepackage{makecell}
\usepackage{utfsym}
\usepackage{verbatim}

\UseRawInputEncoding

\ifCLASSINFOpdf
\else
\fi

\begin{document}
\captionsetup{font={small}}
	%
	\title{Visual Content Privacy Protection: A Survey}
	%
	%
	%
	
\author{Ruoyu~Zhao,\IEEEmembership{}
		Yushu~Zhang,\IEEEmembership{}
		Tao~Wang,\IEEEmembership{}
		Wenying~Wen,\IEEEmembership{}
		Yong~Xiang,\IEEEmembership{}
		and~Xiaochun~Cao\IEEEmembership{}
		\vspace{-1.5em}

\thanks {R. Zhao, Y. Zhang, and T. Wang are with the College of Computer Science and Technology, Nanjing University of Aeronautics and Astronautics, Nanjing	211106, China (e-mail: zhaoruoyu@nuaa.edu.cn; yushu@nuaa.edu.cn; wangtao21@nuaa.edu.cn).}
\thanks {W. Wen  is with the School of Information Technology, Jiangxi University of Finance and Economics, Nanchang 330032, China  (e-mail: wenyingwen@sina.cn).}
\thanks {Y. Xiang is with the School of Information Technology, Deakin University, Burwood, Victoria 3125, Australia (e-mail: yxiang@deakin.edu.au).}
\thanks {X. Cao is with School of Cyber Science and Technology, Sun Yat-sen University, Shenzhen, China (e-mail: caoxch5@sysu.edu.cn).}	}

\markboth{Journal of \LaTeX\ Class Files,~Vol.~xx, No.~xx, September~2023}%
{Shell \MakeLowercase{\textit{et al.}}: Bare Demo of IEEEtran.cls for IEEE Journals}

\maketitle	
\begin{abstract}			
Vision is the most important sense for people, and it is also one of the main ways of cognition. As a result, people tend to utilize visual content to capture and share their life experiences, which greatly facilitates the transfer of information. Meanwhile, it also increases the risk of privacy violations, e.g., an image or video can reveal different kinds of privacy-sensitive information. Researchers have been working continuously to develop targeted privacy protection solutions, and there are several surveys to summarize them from certain perspectives. However, these surveys are either problem-driven, scenario-specific, or technology-specific, making it difficult for them to summarize the existing solutions in a macroscopic way. In this survey, a framework that encompasses various concerns and solutions for visual privacy is proposed, which allows for a macro understanding of privacy concerns from a comprehensive level. It is based on the fact that privacy concerns have corresponding adversaries, and divides privacy protection into three categories, based on computer vision (CV) adversary, based on human vision (HV) adversary, and based on CV \& HV adversary. For each category, we analyze the characteristics of the main approaches to privacy protection, and then systematically review representative solutions. Open challenges and future directions for visual privacy protection are also discussed. 			
\end{abstract} 		

	%
	\IEEEpeerreviewmaketitle

\section{Introduction}

Vision is the most important and complex sense of people, which has become a textbook answer in a sense \cite{10.3389/fpsyg.2019.02246}. For example, there are usually a large number of chapters about vision rather than others in textbooks (e.g., \cite{solso2005cognitive}) on cognitive science. The importance of vision has also been pointed out in some works. In \cite{solso2005cognitive}, it is stated, ``vision is the most widely recognized and the most widely studied perceptual modality''; In \cite{gerrig2015psychology}, it is said, ``vision is the most complex, highly developed, and important sense for humans and most other mobile creatures''. For a reasonable extension, humans prefer to intuitively obtain information from visual content compared with other ways\footnote{https://blog.hubspot.com/marketing/visual-content-marketing-strategy}. An example that illustrates the intuition of visual content can be found in the ancient practice of early humans drawing on rock walls to convey information, and even today, young children learn about the world through picture books. \par

Visual content is better for expressing information than text. this is an era of rough reading; a considerable number of people will not carefully read text, which is a tedious process. For example, research has shown that most of the reading on Twitter occurs only roughly on the first line \cite{2021Visual}. In fact, from a human perspective, the response and processing effect of visual content is better than any other type of data, e.g., the human brain processes images about 60k times faster than text, and about 90\% of the information transmitted to the brain is visual \cite{2021Visual}.\par

People apply visual content data to record and share daily life, important things, and the world, which greatly promotes information exchange and highlights effective information \cite{7488250}. Since the first photo was taken, i.e., Niépce Heliograph was made in 1827 \cite{gernsheim1977150th}, people have entered a new era of recording life with visual content. According to some data, more than 1 trillion photos were taken in 2015 \cite{nightingale2017can}, which is expected to be 1.81 trillion in 2023 \cite{2023ins}. Meanwhile, social networks (OSNs) have occupied most people's lives in the Internet era. Sharing visual content in OSNs is becoming increasingly popular, and people can express themselves and interact with others at any time with a simple click \cite{10.1145/3547299}. According to data, 6.9 billion and 1.3 billion photos are shared on Whatsapp and Instagram every day, respectively \cite{2023ins}. 72 hours of video content will be added to Youtube every minute \cite{7488250}. \par

These captured visual content may reveal privacy information. Traditionally, privacy include but are not limited to personal identity, habits, preferences, and social relations \cite{10.1145/2545883}. Meanwhile, with the continuous development of technology, computer vision can detect more information from visual content (e.g., \cite{10.1145/3558518,9736602,Kim_2022_CVPR}), which has brought great benefits to mankind but also exacerbated the privacy risk. Like the first act of digging a window on the dark wall of the cave, the caveman will be divided into two groups: one group is pleased with the extra sunshine brought by it, and the other group is shocked by the fact that it makes their lives spied \cite{7436680}. In order to obtain sunlight and prevent prying, a ground glass can be added to the window. The purpose of privacy protection is similar to that of ground glass, that is, to balance usability and privacy security in visual content. \par

\subsection{Characteristics of Visual Content Privacy}

In the information age, privacy has received widespread attention due to frequent data leakage, e.g., iCloud leakage event \cite{8264982}, and the introduction of relevant laws and regulations, e.g., EU' General Data Protection Regulation (GDPR) \cite{voigt2017eu}, US' California Consumer Privacy Act \cite{pardau2018california}, and UN' Declaration of Human Rights \cite{assembly1948universal}. Privacy is considered a personal right in a sense, that is, an individual can control the extent to which relevant data are used and disclosed \cite{10.1145/3468877}. If an issue is concerned with the rights of individuals, it is considered a privacy issue. It should be pointed out that the terms `privacy' and `confidentiality' are all response data application guarantees, and they have subtle differences, although they seem to be similar in perception. First, privacy is closely related to a person. Broadly speaking, it refers to the personal right not to be infringed, i.e., the right to protect own personal data is more emphasized \cite{10.2307/2265075}. Second, confidentiality emphasizes that unauthorized persons cannot access data. It emphasizes more on an objective right and a general data protection \cite{10.1145/3468877}.\par

The purpose of privacy protection is that individuals want to prevent data information that is considered privacy sensitive from entering the public domain. When the data are images and videos, this is called \textbf{visual privacy protection} \cite{PADILLALOPEZ20154177}. In this survey, unless otherwise specified, the terms `privacy' and `visual privacy' are the same. \par

\begin{figure*}[htb]
	\setlength{\belowcaptionskip}{-0.5cm} 
	\vspace{-0.4cm}
	\centering 
	\includegraphics[scale=0.45]{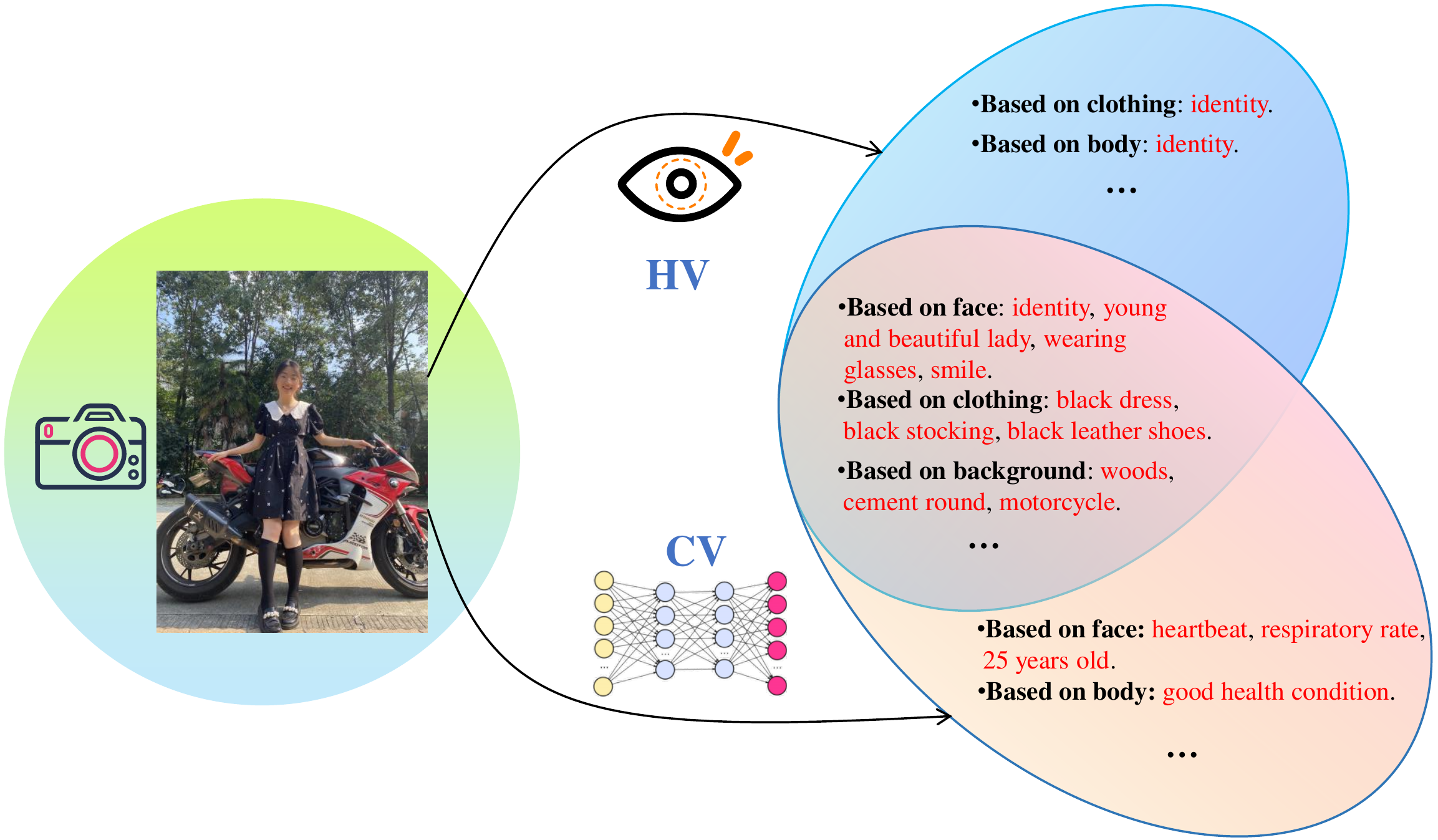} 
	\caption{A simple case of visual content revealing information. For a visual content, CV and HV are able to extract a large amount of information, some of which is unique and some of which has overlap.} 
	\label{fig1}
	\color{blue}
\end{figure*}

Visual content is a rich and subjective source of information as a widely circulated proverb, `\textit{A picture is worth a thousand words}' \cite{HUM20111828}; and similar words can be found in different cultures and language backgrounds. Different readers will extract different meanings from visual content, and readers' reactions will also change with time and environment \cite{chang1997visual}. This makes that using visual content to record information fashionable and will become increasingly popular. However, it also leads to a large amount of privacy information carried by visual content, which is difficult to handle. Meanwhile, even seemingly non-private and mediocre visual content may accidentally reveal privacy.\par

There are three fundamental difficulties in the privacy of visual content, which makes the privacy protection task very challenging.\par

\textbf{The new technology of computer vision (CV) is constantly proposed.} CV is designed to recognize and understand the visual content to extract information. It can bring a lot of benefits but also poses a great threat to privacy. Meanwhile, with the development of CV new technology, the recognition and understanding of visual content by CV have been greatly improved in both depth and breadth. Depth refers to the information that CV is already able to recognize and understand, but it is more accurate with the help of new technology (i.e., fine-grained). Breadth means that information that CV would not have been able to recognize and understand can be done by new technology; i.e., valid information that could not originally be extracted is made by the new technology (diversity). This continuous fine-grained and diversified improvement has broadened the boundaries of privacy caused by CV. In a way, this makes privacy protection and CV a never-ending arms race.\par

\textbf{Human vision (HV) is a sophisticated perception.}  Vision is one of the most important senses for human beings. It is the process of deriving meaning from what is seen, and it is a complex function involving multiple skills. Meanwhile, HV is also not simply a visual input but a combined neurological reasoning process \cite{Kok16275}. For example, a study pointed out that with only a small fragment of the original content, people can also reason about the complete content \cite{10.1145/1978942.1979323}. Despite centuries of research, humans still do not fully understand this process \cite{Hilbert1987-HILCAC}, and HV has been identified as having a subjective nature in a sense \cite{panagiotaropoulos2014subjective}. Thus, although the ability of HV does not seem to have been improving in recent years, new privacy protection schemes for HV are still being proposed.\par

\textbf{Usability requirements for visual content are complex and diverse.} The source of privacy concerns lies in the information carried by visual content. There is no doubt that privacy can be well protected if some methods, e.g., image \cite{9953191} and video \cite{9913480} encryption, are used to discard all information in visual content without distinction \cite{ZHAO2022628}. On the other hand, this also abandons the original purpose of visual content, which is to be used to express information, i.e., usability. Meanwhile, different visual content will have various usability requirements, and even the same visual content in different application scenarios will have different requirements, leading to the complexity and diversion of usability requirements, and the expansion of CV technology in depth and breadth has led to expanding usability requirements.\par

\textbf{Case study 1:} How much information can visual content (for example, an image) reveal?\par

As shown in Fig. \ref{fig1}, this is an image that is often captured in everyday life, and it will be often seen in OSNs, cloud storage platforms, chats, etc. Although it may seem mundane, it can reveal a lot of information. CV and HV can cope well with some direct tasks, such as recognizing identity, gender, beauty, etc, based on the face. On the other hand, CV and HV also have their own unique abilities. The reason for this is that when humans process visual content with a preference for using semantic information, in which CV cannot reliably use, to reason on the whole \cite{NEURIPS2020_9813b270}, while CV prefers low-level information such as texture to process on the data space \cite{5599750}. For those who know the lady in the image, they may be able to identify her simply by her clothes or body outline, which is a high-level and subjective reasoning process and CV is difficult to do \cite{6701391}. For CV, it can be well captured and analyzed some subtle changes beyond HV since it acts in the data space. For example, the heartbeat can cause subtle changes in skin gloss, which can hardly be detected by HV but can be accurately captured by CV. Further, information such as heart rate \cite{Botina-Monsalve_2022_CVPR,9913818} and blood oxygen \cite{10023947} be obtained through analysis. \par

\textbf{Case study 2:} Do privacy/usability requirements differ in different scenarios?\par

For just this one image, there can be complex and even completely opposite privacy and usability requirements. For example, if a user uploads this image to OSNs, chances are she wants to showcase herself to friends but is worried about being probed by CV, e.g., ClearView AI event \cite{10.1145/3446877}, to find out who is in the image. On the other hand, an example is that the image owner wants to apply CV for recognition, but does not want HV to get the information by browsing. For the former, usability is HV to identify people and privacy is blocking CV to identify; For the latter, the opposite is true.\par

It is important to note that this is a simple case, but the actual one is much more complex and diverse than the above. For example, the researchers \cite{HUM20111828} analyzed Facebook profile photos on a tiny scale and came up with a lot of surprising information, e.g., constructing online identity and personality.\par

\subsection{Analysis of Existing Surveys}

There are currently some relevant surveys covering this topic, and they can be divided into three main categories:\par

\begin{itemize}
	\item \textit{Surveys on specific privacy concerns.} They only care about specific concerns without considering the scenarios and technologies used. In \cite{RIBARIC2016131}, privacy concerns arising from identification in multimedia are focused on, and a review summarizes the de-identification mechanism of biometric, non-biometric, and physiologic features in multimedia. In \cite{9481149}, the privacy of face images is concerned, and meanwhile, the main concepts, features, and challenges of protection technologies are introduced, and existing schemes are classified.\par
	
	\item \textit{Surveys on specific scenarios.} They consider privacy concerns in a specific scenario. In \cite{8187822}, the major security and privacy concerns brought by multimedia content sharing OSNs are discussed, and some possible countermeasures are probed. In view of the complex privacy threats faced by image sharing in OSNs, taking the process of image sharing as a clue, the survey \cite{10.1145/3547299} proposes a lifecycle privacy protection framework, and classifies and reviews the existing schemes. The popularity of surveillance cameras has recorded a large amount of visual content with privacy information, and the popularity of CV has led to a rapid increase in the ability to automate the processing of the information, which in turn poses a huge threat to privacy. In \cite{PADILLALOPEZ20154177}, a comprehensive review of the image privacy is presented for surveillance scenarios, and classification and summary of schemes are presented. The deployment of automated face detection/recognition systems has been a significant threat to personal privacy In \cite{10.1145/3583135}, a summary survey of works addressing such concerns is presented, discussing their limitations and future possibilities.\par
	
	\item \textit{Surveys on specific technologies.} They consider specific technologies to deal with some privacy concerns. In \cite{10.1145/3436755}, privacy attacks, as well as protections based on machine learning, are reviewed, which deals with attacks on and protection of visual content. In \cite{10.1145/3490237}, it is reviewed for protection schemes for differential privacy applications to unstructured content, which involves (but are not limited to) image and video data.
\end{itemize}

\begin{table*}[htb]
	\centering
	\caption{Comparison of this survey with existing surveys.}
	\renewcommand{\arraystretch}{1.5}
	\tabcolsep 10pt
	
	\label{tab: compare}
	\begin{tabular}{cccccccc}
		\toprule[2pt]
		\multicolumn{1}{c}{Paper} & Year & \begin{tabular}[c]{@{}c@{}}Adversary\\ CV\end{tabular} & \begin{tabular}[c]{@{}c@{}}Adversary\\ HV\end{tabular} & Image & Video & \begin{tabular}[c]{@{}c@{}}Non-biometric\\ feature\end{tabular} & Focus \\ \hline
		\cite{RIBARIC2016131} & 2016 & \CIRCLE & \CIRCLE &  \CIRCLE     &  \CIRCLE     & \LEFTcircle &  De-identification \\
		
		\cite{9481149}	& 2021 & \CIRCLE & \Circle & \CIRCLE & \Circle & \Circle &  Face biometrics \\
		
		\cite{8187822} & 2015 & N/A & N/A & \CIRCLE & \CIRCLE &  N/A & \makecell{Cyber and system\\ security in OSNs} \\
		
		\cite{10.1145/3547299} & 2023 & \CIRCLE & \CIRCLE & \CIRCLE & \Circle & \CIRCLE & \makecell{Image content\\ in OSNs} \\
		
		\cite{PADILLALOPEZ20154177} & 2015 & \CIRCLE & \CIRCLE & \CIRCLE & \Circle & \LEFTcircle &  \makecell{Image content\\ in surveillance} \\
		
		\cite{10.1145/3583135}	& 2023 & \CIRCLE & \Circle & \CIRCLE & \CIRCLE & \Circle & \makecell{Biometric facial\\  recognition system}   \\
		
		\cite{10.1145/3436755} & 2021 & \CIRCLE & \Circle & \CIRCLE & \CIRCLE & N/A & \makecell{Machine learning\\ privacy} \\
		
		\cite{10.1145/3490237}	& 2022 & N/A & N/A & \CIRCLE & \CIRCLE & N/A &  \makecell{Differential privacy\\  for unstructured data} \\
		
		This survey & - & \CIRCLE & \CIRCLE & \CIRCLE & \CIRCLE & \CIRCLE &  \makecell{Macro understanding of\\ visual content privacy} \\
		
		\bottomrule[2pt]
		
	\end{tabular}
	
\end{table*}

A quantitative comparison of this survey with the above surveys is shown in Table \ref{tab: compare}, in which the meaning of symbols are as follows:

\begin{itemize}
	\item \CIRCLE: The survey explicitly mentions this item;
	\item \LEFTcircle: Although some (very small) percentages in the surveys deal with non-biometric features, their goal is still to protect biometric ones;
	\item \Circle: The survey explicitly does not include this item;
	\item N/A: No relevant items are mentioned.
\end{itemize}

In all, previous surveys often focused on a specific privacy (e.g., de-identification \cite{RIBARIC2016131}) or on a specific scenario (e.g., OSNs \cite{10.1145/3547299}) or on the impact that a particular technology may have on the privacy protection of visual content (e.g., machine learning \cite{10.1145/3436755}). They achieve the intended goal well but are inadequate in terms of privacy protection of visual content. For example, de-identification is indeed an important topic in privacy protection, but concerns in visual content go far beyond personal identifiers (e.g., health status); OSNs are indeed one of the significant streams of visual content, and also generate a great of privacy concerns as a result (e.g., ClearView AI event \cite{10.1145/3446877}), but the application scenarios for visual content are not limited to OSNs either (e.g., cloud storage); machine learning is an important source of privacy threats to visual content and an influential enabler of privacy protection, but HV is also a major threat to privacy, and non-machine learning methods can also be good for privacy protection (e.g., blur and mosaic \cite{PADILLALOPEZ20154177}).\par

The fundamental difference between this survey and previous pioneering works is that it contains all privacy concerns and countermeasures related to visual content, rather than a limited set of preferences. The key differences are as follows:
\begin{itemize}
	\item Previous surveys have often been conducted from a problem-driven perspective, i.e., presupposing a specific concern or scenario. Instead, this survey tries to build a more comprehensive overview of the field by focusing on the visual content itself.
	
	\item Previous surveys often focused on the privacy concerns posed by CV, and even when HV is considered, they do not place CV and HV at the same level of narrative on an equal footing. In contrary, this survey considers the threat to privacy posed by CV and HV to be equally important in terms of for visual content.
	
	\item The privacy threats considered in previous surveys are mainly brought by biometrics, and they do not consider non-biometrics, or consider them also for de-biometrics. Conversely, in this survey, non-biometric features are also an important part of privacy protection, especially for HV.
\end{itemize}

In brief, this survey provides a comprehensive understanding of visual content privacy through the following contributions: 1) the characteristics and taxonomy of visual content privacy; 2) a novel privacy protection framework for visual content; 3) any visual privacy protection scheme can find its place in this framework; 4) any privacy concerns regarding visual content can find potential solutions here; 5) an in-depth review of current advances in visual privacy protection solutions; 6) a discussion of the challenges and future directions of visual privacy protection.\par

The remainder of this survey is organized as follows. Section \ref{Section: overview} presents a taxonomy based on adversaries, along with a description of each adversary and their relationships and differences. Sections \ref{Section: CV adversary} to \ref{Section: mixed adversary} are specific to each category, review representative solutions, and discuss common principles. Section \ref{Section: challenges} discusses the challenges and future directions of visual privacy protection. Section \ref{Section: conclusion} is a brief conclusion.\par

\begin{figure*}
	\centering 
	\subfigure[]{\includegraphics[scale=0.17]{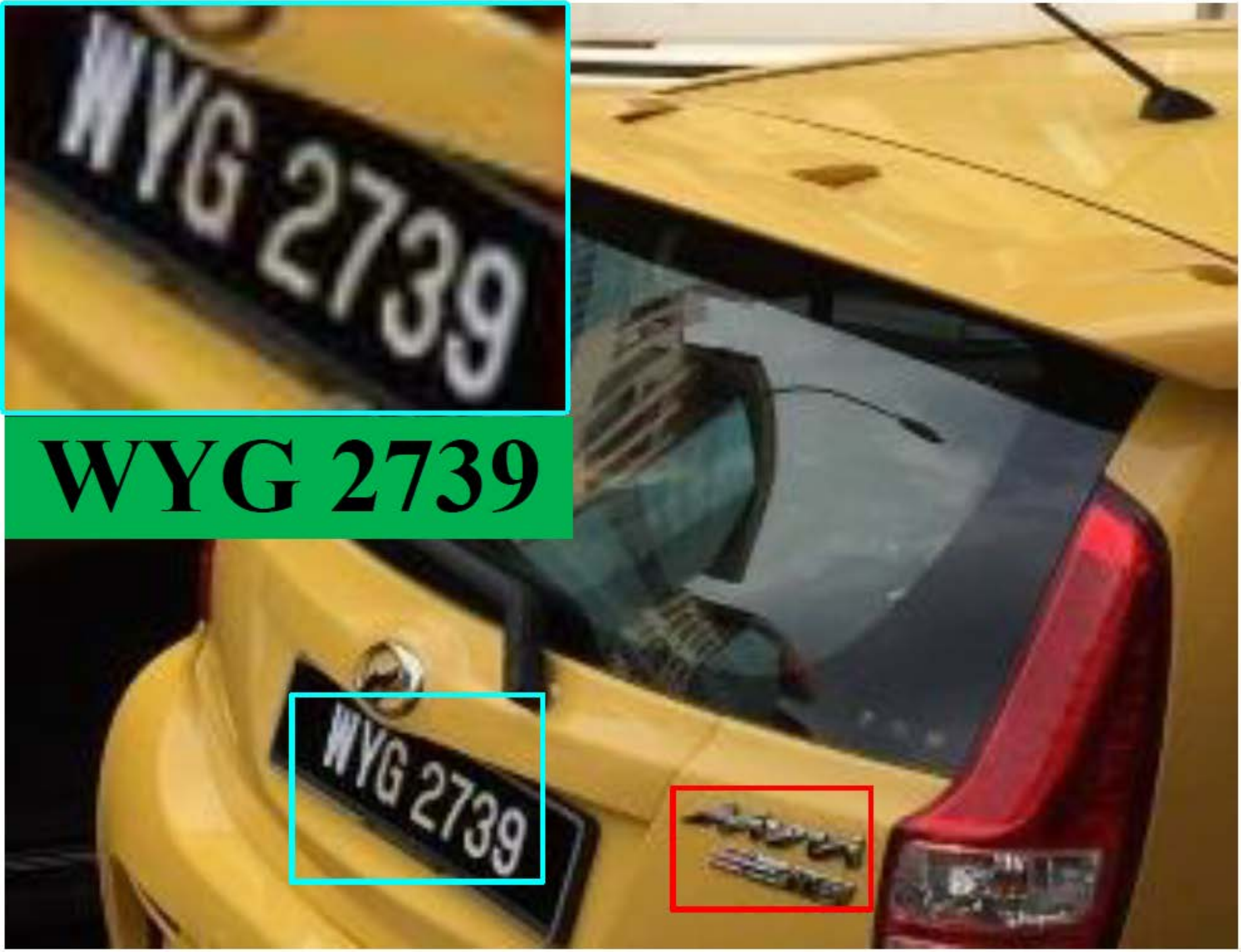}\label{text1}}\vspace{-0.0em}
	\subfigure[]{\includegraphics[scale=0.17]{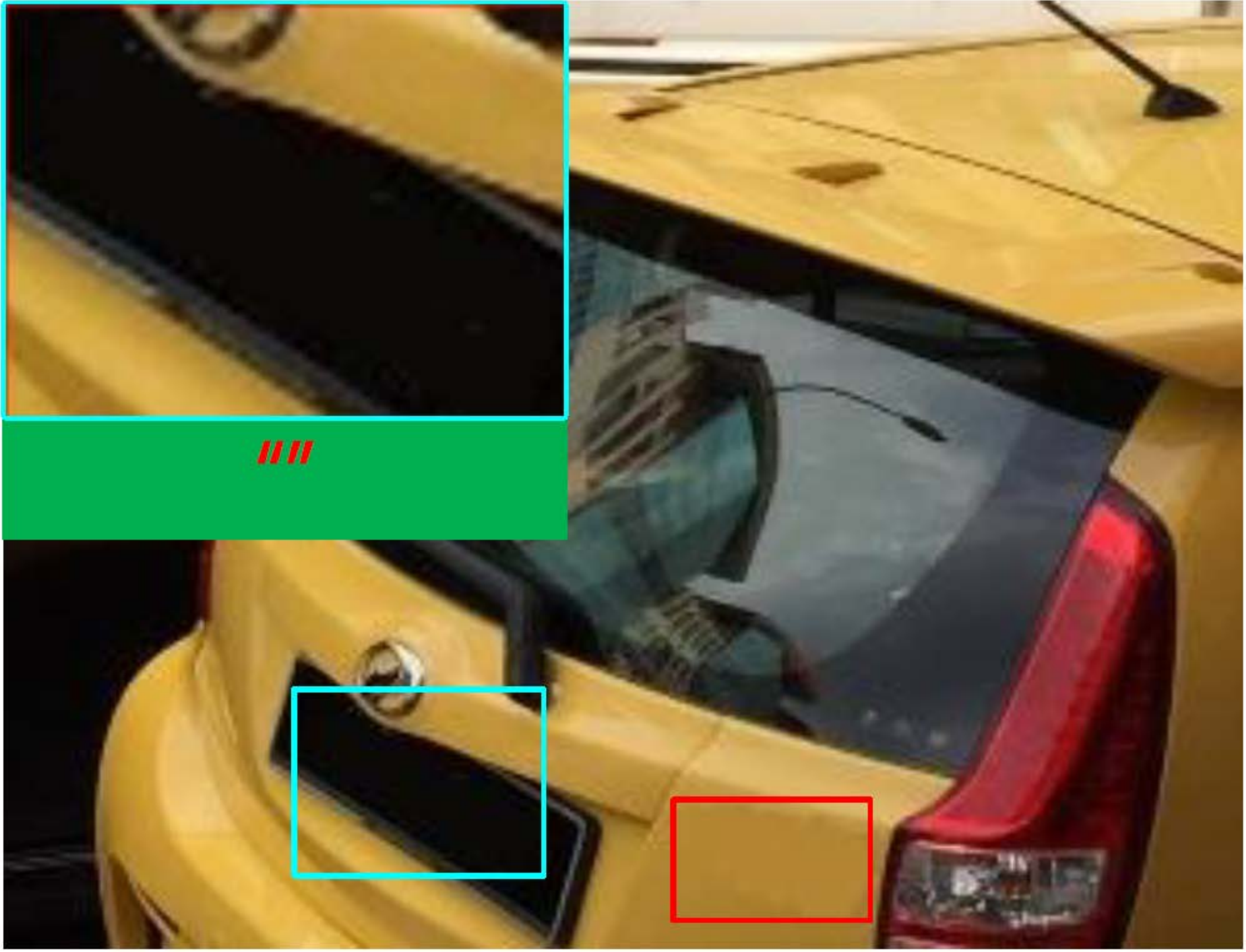}\label{text2}}
	\subfigure[]{\includegraphics[scale=0.17]{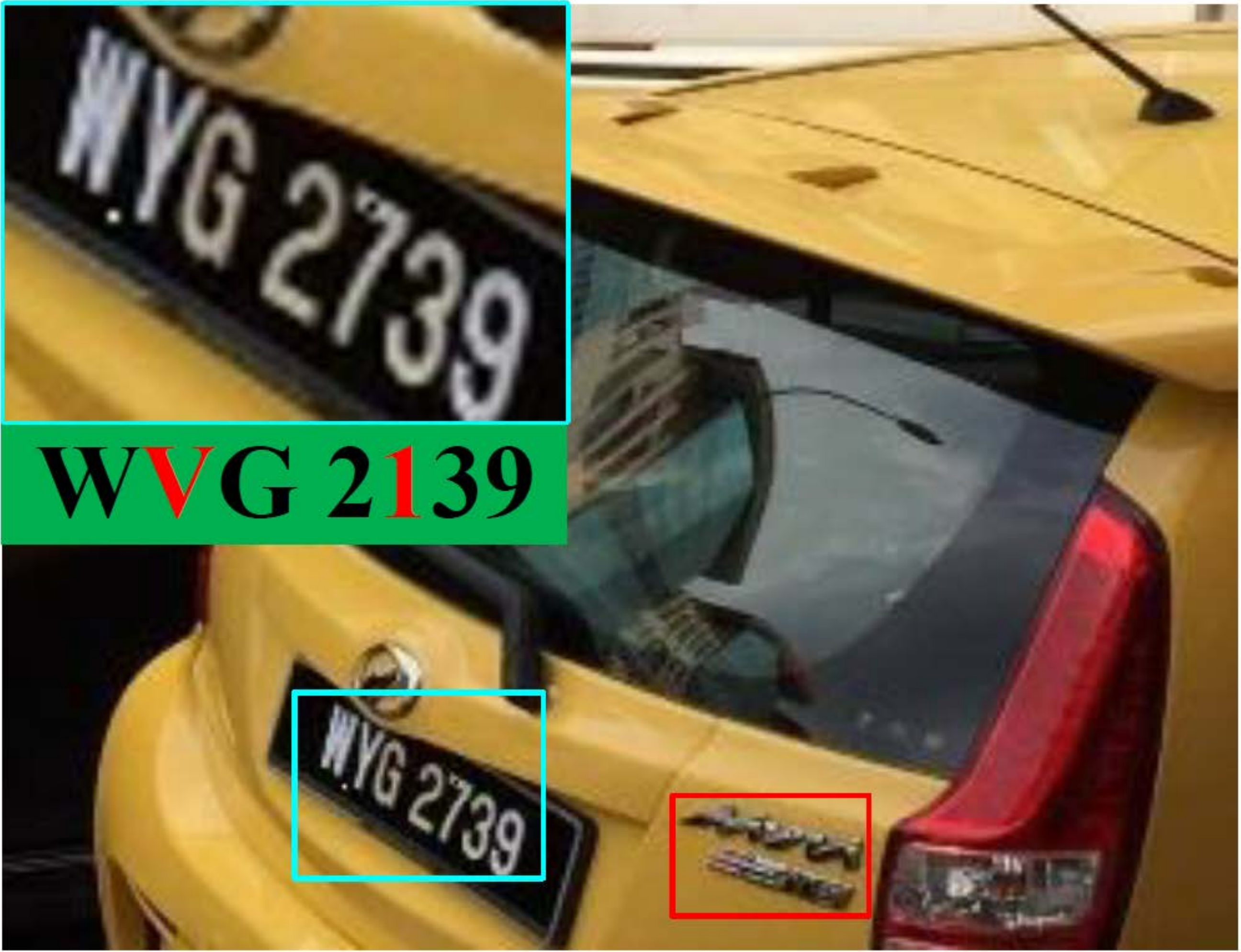}\label{text3}}
	
	\caption{Examples (from \cite{10045728}) against HV adversaries VS. against CV adversaries, in which text is the privacy content. (a) original image; (b) against HV adversaries; (c) against CV adversaries. As shown in Fig. \ref{text2}, if the protected content (e.g., license plate number area) has a feature that can be detected by a specific CV for the text `WYG 2739', then it is said to be protected against HV\&CV adversaries.}
	\label{text}
\end{figure*}

\begin{figure*}[htb]
	\setlength{\belowcaptionskip}{-0.5cm} 
	\centering 
	\includegraphics[scale=0.45]{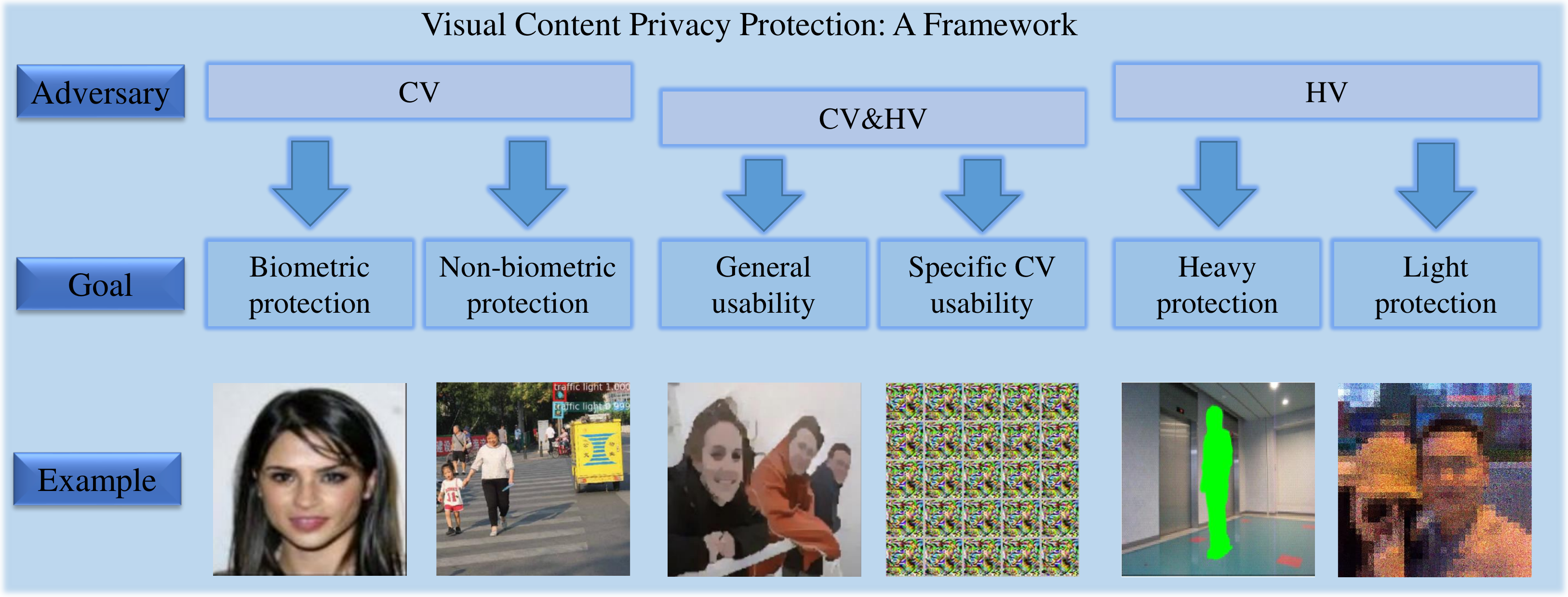} 
	\caption{Visual content privacy protection framework, including three adversaries: CV, HV, and CV\&HV. Examples from left to right are from, respectively, \cite{Maximov_2020_CVPR}, and \cite{9148656}, \cite{10.1145/3447993.3448618}, \cite{10.1145/3524273.3528189}, \cite{10.1007/978-3-540-77409-9_14}, and \cite{9393403}.} 
	\label{framework}
\end{figure*}

\section{Visual Content Privacy: An Overview} \label{Section: overview}

\subsection{Adversary-Based Taxonomy}

The field of information security is essentially an offensive and defensive confrontation, proposing the targeted defense solution for a specific security threat (i.e., adversary). In other words, the role of an adversary can be found in any given work. Privacy protection, in a sense, also belongs to the field of information security, and any specific scheme can also find a privacy adversary. The taxonomy by adversaries can include all privacy threats and protection solutions for visual content. There are three adversaries of privacy for visual content.\par

\textbf{CV adversary:} It refers to an adversary that uses only artificial intelligence to reason or associate sensitive information implied in visual content. Essentially, CV is an attempt to replicate the HV system in order to understand visual content. Therefore, it can directly recognize some observable information of visual content, such as face identity, color, license plate number, etc., just like HV. Meanwhile, for certain information that is difficult for humans to perceive, such as soft biometric attributes \cite{9770950}, CV is good at it which can be accurately reasoned in latent feature space.\par

\textbf{HV adversary:} It refers to an adversary that uses only the human eyes to directly view visual content to obtain sensitive information such as the face, signature, license plate number, etc. This is the most intuitive privacy adversary, since the original purpose of visual content is to convey information to people through browsing. This is also the easiest adversary to achieve privacy breaches since once the visual content is accessed maliciously, the HV adversary will immediately gain access to privacy-sensitive information.\par

\textbf{CV\&HV adversary:} It refers to an adversary that uses both CV and HV to infringe on sensitive information in visual content. As shown in Fig. \ref{fig1}, some information can be recognized and extracted by both CV and HV, such as face identity, gender, and license plate number, etc. The difference between this adversary and the above two is that the above only considers pure CV and HV threats to privacy. However, this part considers a combination of the both privacy threats and takes into account the usability of a specific CV.\par

The differences and connections between these three are as follows:

\begin{itemize}
	\item Protection against CV adversaries considers only attacks on CV, with visual content changes as small as possible, with no or minimal impact on HV. This is because CV is far less robust than HV and only tiny changes are required to achieve protection against CV.
	
	\item Protection against HV tends to work for CV, and the opposite does not hold \cite{9481149}. The protection of HV should significantly change the visual content since HV is robust and tends to learn the visual content in its entirety. An example is shown in Fig. \ref{text}. One of the protection against HV adversaries is to erase all text, such as license plate numbers, which naturally also has a decisive effect on CV. The protection against HV adversaries is added with noise that has no effect on the text extracted by HV.
	
	\item If the solution for HV adversaries considers the usability of the specific CV, then CV is divided into usability CV and adversary CV, and it becomes a protection against HV\&CV adversary. Specifically, the difference between protection against HV\&CV and against HV only is that the former takes into account the usability of the protected content to a specific CV (model or task), while the latter does not. As shown in Fig. \ref{text2}, an example is given: if the protected content (e.g., license plate number area) has a feature that can be detected by a specific CV for the text `WYG 2739', then it is said to be protected against HV\&CV.
	
	\item In all, the ranking of the privacy protection ability of solutions is HV\&CV $\textgreater$ HV $\textgreater$ CV. Specifically, privacy protection schemes for HV\&CV adversaries also work for CV adversaries; and for CV adversaries also work for HV ones. However, the opposite is not valid.
	
\end{itemize}

\subsection{Privacy Protection Framework}

We design an adversary-based privacy protection framework to help identify different privacy concerns in visual content and investigate corresponding countermeasures. The framework covers all privacy adversaries that visual content can may encounter, forming a closed loop for privacy analysis, and any scheme of visual privacy protection can find a suitable classification in this framework, as shown in Fig. \ref{framework}.\par

\subsubsection{CV adversary}
\begin{itemize}
	\item \textbf{Biometric protection:} It prevents CV from extracting biometric features from visual content (mainly faces), which is one of the main directions of interest to the privacy community today.\par
	\item \textbf{Non-biometric protection:} It prevents CV from learning about non-biological features from the visual content, e.g., license plate number and location.\par
\end{itemize}

\subsubsection{HV adversary}

\begin{itemize}
	\item \textbf{Local heavy protection:} It targets sensitive areas in visual content and completely eliminates the visual effects therein. \par
	\item \textbf{Global light protection:} Its goal is to eliminate all the refined visual content, so that the naked eye can only observe the rough content.\par	
\end{itemize}

\subsubsection{CV\&HV adversary}

\begin{itemize}
	\item \textbf{Partial usability preserved:} It prevents CV and HV from identifying specific or generalized private information, while preserving certain visuals for CV and HV to perform cursory tasks, e.g., counting. \par
	\item \textbf{Specific CV usability:} It blocks CV and HV from identifying private information while not preserving any visuals, but allows specific CV models to perform a limited task.\par	
\end{itemize}

\begin{figure*}[htb]
	\setlength{\belowcaptionskip}{-0.5cm} 
	\centering 
	\includegraphics[scale=0.45]{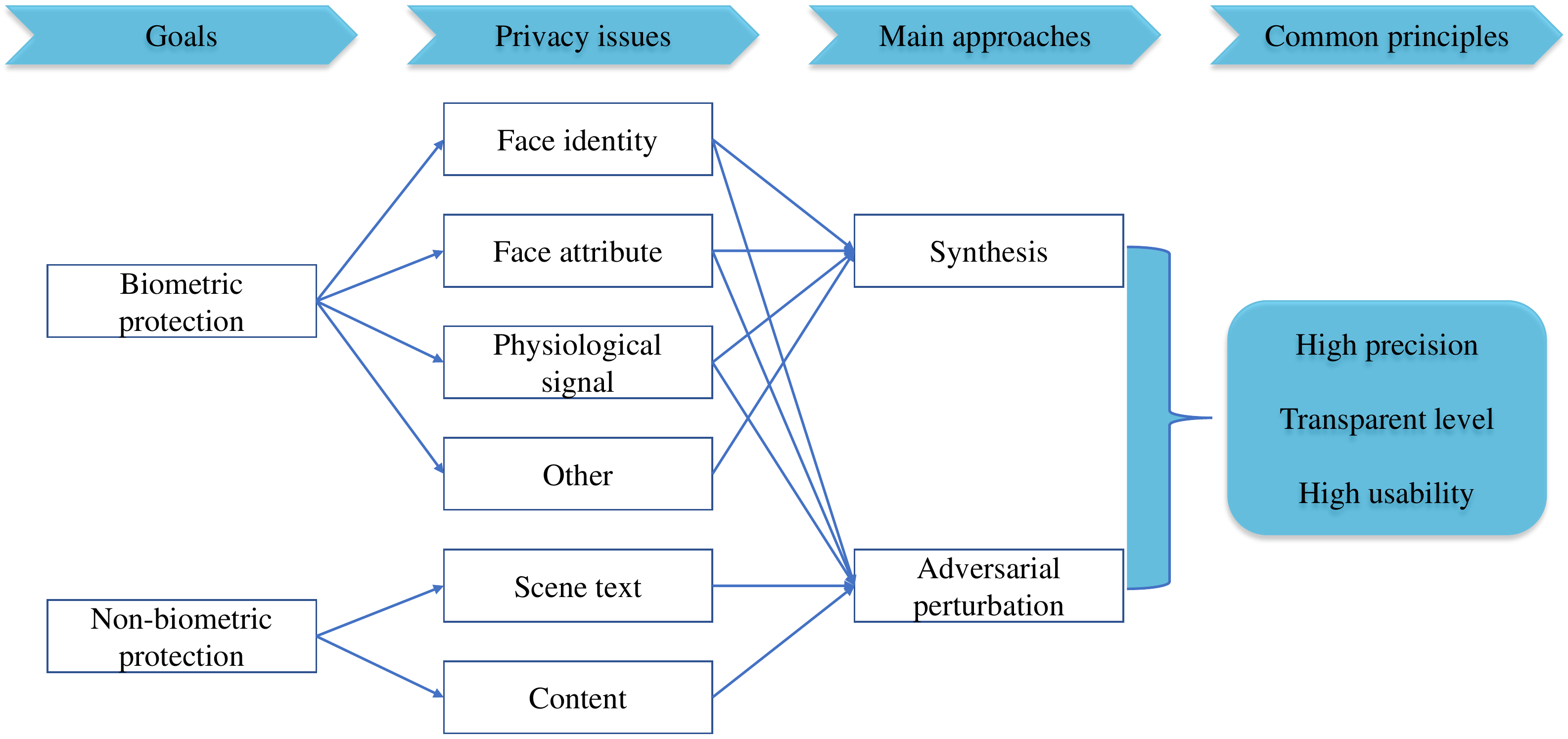} 
	\caption{Overview of the privacy protection analysis for CV adversary.} 
	\label{CV adversary framework}
\end{figure*}

\section{Privacy Protection for CV Adversary} \label{Section: CV adversary}

This section provides a comprehensive analysis of privacy concerns posed by CV-based adversaries to visual content, including privacy issues, main approaches, and common principles, as shown in Fig. \ref{CV adversary framework}.\par

\subsection{Privacy Issues in CV Adversary}

\subsubsection{Issues associated with biometric protection}

Extracting biometrics from visual content, especially face, using CV has always been an important interest in the field of artificial intelligence due to its practicality. A rich variety of biometric features are extracted from the visual content represented by face, and the accuracy and variety continue to expand, have also led to serious privacy concerns. Depending on the focus of biometrics, the corresponding protection can be divided into four categories.

\begin{itemize}
	\item \textbf{Face identity:} One of the major purposes of taking photos or videos is to show oneself, which inevitably records the person's face. One of the most mature, highly regarded, and widely used technologies in the CV field is facial identity recognition \cite{8614364}. Meanwhile, face identity is also one of the hardest hit areas for privacy issues from visual content,  e.g., ClearView AI event \cite{10.1145/3446877}, and accordingly one of the focuses of research in the privacy community.
	\item \textbf{Face attribute:} As CV's ability to analyze and process visual content has increased, CV' analysis of faces has expanded beyond identity to include attribute features such as age, expression, race, and gender. These attributes may help the adversary to spy on people more accurately, and in a sense, attributes are more sensitive than identity, which can give rise to bias and discrimination (e.g., race and gender) \cite{Dhar_2021_ICCV}.
	\item \textbf{Physiological signal:} In the video content, not only the information about a person's identity and attributes exist, but also the physiological signals of the person are captured \cite{9680677}. The study points to the feasibility of remote photoplethysmography (rPPG) based on visual content via CV, e.g., heart rate \cite{9220114} and respiration rate \cite{Du_2021_ICCV}. These signals may be secretly analyzed and used by intentional people, which may pose a threat to privacy, such as gaining advantages in negotiation and analyzing health status.
	\item \textbf{Other:} The studied described above are biometric protections of great concern to the privacy community since they pose intuitive privacy concerns that ordinary people are often aware of in their daily lives. On the other hand, CV is also capable of extracting other biometrics that pose privacy threats in addition to the types mentioned above, e.g., gait \cite{Zhu_2021_ICCV} and eye gaze \cite{9028119}, and there is some sporadic but equally important work that considers this privacy.
\end{itemize}

\subsubsection{Issues associated with non-biometric protection}

CV work on the non-biometric content is not as focused as biometric features (often for faces), focusing mainly on scene text and content analysis. 

\begin{itemize}
	\item \textbf{Scene text:} Text is one of the basic media for transmitting information, and it is ubiquitous in daily life, such as street nameplates and license plate numbers. This kind of text in the natural environment is called scene text \cite{Wang_2019_CVPR}, which is one of the focuses of CV attention. On the other hand, scene text can contain all kinds of sensitive information (e.g., personal names and addresses), and with the advancement of CV, the resulting privacy risks are greater.
	
	\item \textbf{Content:} CV is able to analyze visual content and infer the presence of sensitive non-biometric information such as scene and clothing, which can also adversely affect privacy if used by adversaries \cite{Orekondy_2018_CVPR}.\par
\end{itemize}

\subsection{Main Approaches} \label{Section:3-2}

\textbf{Synthesis:} It refers to privacy protection by replacing the sensitive content in the original visual content with the privacy-preserving one that can be pre-existing or generated. With the advent of deep generative models, such as Generative Adversarial Networks (GAN), great progress has been made in building high quality generative models. For face privacy protection, privacy-sensitive features are usually extracted from the original face, modified (e.g., eliminated or obfuscated), and then the protected features are combined with other necessary features are fed together into a generator to reconstruct the privacy-preserving face. \par

\textbf{Adversarial perturbation:}  Research has shown that CV can perceive features that are not perceived by humans but are critical to CV to recognize objectives \cite{NEURIPS2019_e2c420d9}, making CV particularly sensitive to some perturbations that are not perceptible to humans. Therefore, corruption of these features, i.e., using carefully crafted perturbations added directly to the original visual content, can effectively prevent CV from recognizing the according sensitive features by CV.\par

\subsection{Solutions for Biometric Protection}

\subsubsection{Face identity}\ 

\textit{Synthesis.} Bitouk \textit{et al.} \cite{10.1145/1399504.1360638} created a large face database by crawling web images, protecting identities by comparing the face in the source image with faces in the database and selecting a face with the highest scores to swap. Afterward, the swapped faces are re-illuminated and recolored to blend into the image as much as possible and produce visually convincing results. Mosaddegh \textit{et al.} \cite{10.1007/978-3-319-16811-1_11} achieved identity protection by dividing a face into some components and then using the different components from multiple faces in the database for corresponding swaps. It somehow ensures that the protected faces do not resemble any human face and protects the privacy of the donor. 2D face swapping requires the face in the source image to have similar pose and appearance to the face in the target one, which affects its application. Lin \textit{et al.} \cite{6298419} proposed a 3D model-based face swapping scheme that is capable of rendering any pose of the head, eliminating restrictions on pose and appearance similarity, and enabling the swapping of front or side faces.\par

GAN has been shown to be successful in generating realistic face regions and is therefore heavily used in existing schemes to protect face privacy \cite{Nirkin_2019_ICCV}. DeepFake \cite{2027deepfakes}, a deep forgery technique using GAN, is proposed for face swapping, which is robust to different facial expressions, lighting, and poses, and is also applicable to video. Zhu \textit{et al.} \cite{10.1145/3375627.3375849} proposed the use of DeepFake applied to medical videos to protect the privacy of patient's identity while applying to the medical examination of Parkinson. Nirkin \textit{et al.} \cite{Nirkin_2019_ICCV} proposed a face swap scheme based on GAN that can manipulate pose, expression, and change identity simultaneously.  \par

Face swapping schemes protect identity privacy in the source content by transferring the face from the target content to the source one. However, there are some problems. First, this often requires a target face or even a large database of faces. Second, although the identity of the source face is protected, the information of the target one is compromised.\par

Some GAN-based identity protection schemes that directly suppress or eliminate the original identity features of faces have also been proposed. Maximov  \textit{et al.} \cite{Maximov_2020_CVPR} proposed a framework for face anonymization based Conditional GAN for images and videos, where faces are anonymized based on Siamese networks to provide guidance for identity signals. Wen \textit{et al.} \cite{WEN2022197} combined differential privacy and GAN to propose an image anonymization scheme that can adjust the balance of privacy and usability through privacy budget parameters, while explicitly proposing to maintain the attributes unchanged. Zhai \textit{et al.} \cite{10.1145/3503161.3547757} pointed out that considering anonymization purely in terms of identity characteristics may lead to uncontrollable changes in some attributes. Therefore, they proposed a scheme to change identity by controlling changes in attributes, so as to achieve the three requirements of anonymity, realism, and controllability. In addition, Gu \textit{et al.} \cite{10.1007/978-3-030-58592-1_43} combined passwords with GAN to propose reversible face anonymization schemes that recover the original face when the user gives the correct password and generate faces with different identities when different passwords are used.\par

The GAN-based scheme to synthesize facial content is indeed good at ensuring the quality of the generated visual content, while it does not preserve the face of the original face, and thus not only prevents CV from recognizing identity, but may also mislead HV (although it is not intended to do so).\par

\textit{Adversarial perturbation.} 
Chatzikyriakidis \textit{et al.} \cite{8803803} proposed the use of penalized fast gradient value method to generate adversarial perturbations to achieve face de-identification, which is guaranteed to be visually very similar to the original image. Cherepanova \textit{et al.} \cite{cherepanova2021lowkey} addressed the privacy problem of capturing images on OSNs to build face databases, and proposed the first tool to block commercial face API to evade face recognition by adding adversarial perturbation before uploading images to OSNs. Zhong \textit{et al.} \cite{9778974} proposed a custom cloak face privacy protection using adversarial perturbation, in which each identity has a personalized cloak and all images about this identity can apply the clock.\par

The face protection scheme using adversarial perturbation ensures that the information perceived visually by humans is consistent with the original visual content. However, it is often computationally expensive and needs to be retrained for each face or each identity to gain perturbations.\par

\begin{table*}[htb]
	\centering
	\caption{A summary of the solutions in biometric protection.}
	\renewcommand{\arraystretch}{1.5}
	\tabcolsep 7pt
	\label{tab: cv-b}
	\begin{tabular}{cccccccccc}
		\toprule[2pt]
		
		\multirow{2}{*}{\makecell{Sub\\class}}            & \multirow{2}{*}{Paper} & \multirow{2}{*}{Year} & \multirow{2}{*}{\makecell{Key \\technology}} & \multicolumn{2}{c}{HV} & \multirow{2}{*}{\makecell{Reversi-\\bility}} & \multicolumn{2}{c}{Type} & \multirow{2}{*}{Note} \\ \cline{5-6} \cline{8-9}
		&                        &                       &                                 & Reality  & Invariance &                                & I       & V      &                       \\ \hline
		\multirow{12}{*}{ \makecell{Face\\ identity}}       & \cite{10.1145/1399504.1360638} & 2008 & \makecell{face\\ swapping} & \usym{2610} & \usym{2612} & \usym{2612} & \usym{2611} & \usym{2612} &  \makecell{Infringed the\\ identity of others}                \\
		& \cite{10.1007/978-3-319-16811-1_11} & 2015 & \makecell{face\\ swapping} & \usym{2610} & \usym{2612} & \usym{2612} & \usym{2611} & \usym{2612} & \makecell{Multi-facial\\ component splicing} \\
		& \cite{6298419} & 2012 & \makecell{face\\ swapping} & \usym{2610} & \usym{2612} & \usym{2612} & \usym{2611} & \usym{2612} & \makecell{Arbitrary\\ face pose} \\
		& \cite{10.1145/3375627.3375849} & 2020 & DeepFake & \usym{2611} & \usym{2612} & \usym{2612} & \usym{2612}  &  \usym{2611} & \makecell{Medical\\ applications} \\
		& \cite{Nirkin_2019_ICCV} & 2019 & GAN & \usym{2611} & \usym{2612} & \usym{2612}& \usym{2611} & \usym{2611} & \makecell{Arbitrary\\ face pose} \\
		& \cite{Maximov_2020_CVPR} & 2020 & GAN & \usym{2611} & \usym{2612} & \usym{2612} & \usym{2611} & \usym{2611} & \makecell{Identity\\ vector guidance} \\
		& \cite{WEN2022197} & 2022 & GAN & \usym{2611} & \usym{2612} &  \usym{2612} & \usym{2611} & \usym{2612} & \makecell{Differential\\ privacy}   \\
		& \cite{10.1145/3503161.3547757} & 2022 & GAN & \usym{2611} & \usym{2612} & \usym{2612} & \usym{2611} & \usym{2612} & \makecell{Control attributes\\ to change identity} \\
		& \cite{10.1007/978-3-030-58592-1_43} & 2020 & GAN & \usym{2611} & \usym{2612} & \usym{2611} & \usym{2611} & \usym{2612} & \makecell{New identity used\\ for wrong password} \\
		& \cite{8803803} & 2019 &  Perturbation  & \usym{2611} & \usym{2611} & \usym{2612} & \usym{2611} & \usym{2612} & - \\
		& \cite{cherepanova2021lowkey} & 2021 & Perturbation & \usym{2611} & \usym{2611} &  \usym{2612} & \usym{2611} & \usym{2612} &  \makecell{Against\\ commercial APIs} \\
		& \cite{9778974} & 2023 & Perturbation & \usym{2611} &  \usym{2611} & \usym{2612} & \usym{2611} & \usym{2611} &  \makecell{One identity\\ one cloak} \\ \hline
		\multirow{6}{*}{\makecell{Face\\ attribute}}       & \cite{6247759} & 2012 & \makecell{3D face\\ reconstruction}   & \usym{2611} &  \usym{2612} &  \usym{2612}  & \usym{2612} & \usym{2611} &  \makecell{Focused on\\ expression} \\
		& \cite{8718508} & 2019 & GAN & \usym{2611} & \usym{2612} &  \usym{2612} &  \usym{2611} & \usym{2612} &  \makecell{Only change\\ target attributes}  \\
		&  \cite{Wang_2021_CVPR} & 2021 & GAN & \usym{2611} & \usym{2612} & \usym{2612} & \usym{2611} & \usym{2612}  & \makecell{Attribute\\ obfuscation} \\
		& \cite{9201364} & 2020 & GAN & \usym{2611} & \usym{2612} & \usym{2612} &  \usym{2611} & \usym{2612} & \makecell{Multi-\\ attribute} \\
		& \cite{8272743} & 2017 & Perturbation & \usym{2612} & \usym{2611} &  \usym{2612} & \usym{2611} & \usym{2612} & \makecell{Focused on\\ gender} \\
		& \cite{9746848} & 2022 & Perturbation & \usym{2611} & \usym{2611} & \usym{2612}  & \usym{2612} & \usym{2611} & \makecell{Focused on\\ emotion} \\ \hline
		\multirow{3}{*}{\makecell{Physio-\\logical\\ signal}} & \cite{7961722} & 2017  & Elimination & \usym{2610} & \usym{2611} & \usym{2612} & \usym{2612} & \usym{2611} & \makecell{Unrealistic videos\\ may appear} \\
		& \cite{9680677} & 2022  & Perturbation & \usym{2611} &  \usym{2611} &  \usym{2612}  &  \usym{2612} & \usym{2611} &  \makecell{Designating\\ rPPG}  \\
		& \cite{9806161} & 2022 & Deep learning  & \usym{2611} &  \usym{2611} &  \usym{2612}  &  \usym{2612} & \usym{2611} &  Real time   \\ \hline
		\multirow{3}{*}{Other}                & \cite{7350779} & 2015 & Reshaping &  \usym{2611} & \usym{2612} & \usym{2612}  & \usym{2612} &  \usym{2611} & Preserving posture \\
		&  \cite{10.1145/3503161.3548235} & 2022 & GAN & \usym{2611} &  \usym{2612} & \usym{2612} & \usym{2611} &  \usym{2612} &  \makecell{pedestrain\\ anonymination} \\
		
		\bottomrule[2pt]
	\end{tabular}
\end{table*}

\begin{table*}[htb]
	\centering
	\caption{A summary of the solutions in non-biometric protection.}
	\renewcommand{\arraystretch}{1.5}
	\tabcolsep 7pt
	\label{tab: cv-nb}
	\begin{tabular}{cccccccccc}
		\toprule[2pt]
		
		\multirow{2}{*}{\makecell{Sub\\class}}            & \multirow{2}{*}{Paper} & \multirow{2}{*}{Year} & \multirow{2}{*}{\makecell{Key \\technology}} & \multicolumn{2}{c}{HV} & \multirow{2}{*}{\makecell{Reversi-\\bility}} & \multicolumn{2}{c}{Type} & \multirow{2}{*}{Note} \\ \cline{5-6} \cline{8-9}
		&                        &                       &                                 & Reality  & Invariance &                                & I       & V      &                       \\ \hline
		
		\multirow{4}{*}{ \makecell{Scene\\ text}}  
		
		& \cite{9162685} & 2020 & Perturbation & \usym{2611} & \usym{2611} & \usym{2612} & \usym{2611}  &  \usym{2612} & \makecell{White-box} \\
		
		& \cite{10.1007/978-3-030-67664-3_33} & 2021 & Perturbation & \usym{2611} & \usym{2612} & \usym{2612}& \usym{2611} & \usym{2612} & \makecell{White-box} \\

		& \cite{Xu_2020_CVPR} & 2020 & Perturbation & \usym{2611} & \usym{2611} & \usym{2612} & \usym{2611}  &  \usym{2612} & \makecell{White-box} \\
		
		& \cite{10045728} & 2023 & Perturbation & \usym{2611} & \usym{2611} & \usym{2612}& \usym{2611} & \usym{2611} & \makecell{Black-box} \\
		
		\hline
		
		\multirow{5}{*}{Content} & \cite{liu2017protecting} & 2017  & Perturbation & \usym{2611} & \usym{2611} & \usym{2612} & \usym{2611} & \usym{2612} & \makecell{Content all\\ protected} \\
		
		& \cite{9806161} & 2020  & Perturbation & \usym{2611} &  \usym{2611} &  \usym{2612}  &  \usym{2611} & \usym{2612} &  \makecell{Optional protected\\ content}  \\
		
		& \cite{Duan_2021_ICCV} & 2021 & Perturbation  & \usym{2611} &  \usym{2611} &  \usym{2612}  &  \usym{2611} & \usym{2612} &  \makecell{Dropping\\ information}   \\
		
		& \cite{10.1145/3343031.3351088} & 2019 & Perturbation &  \usym{2611} & \usym{2611} & \usym{2612}  & \usym{2612} &  \usym{2611} & Black-box \\
		
		& \cite{rajabi2021practicality} & 2021 & Perturbation &  \usym{2612} & \usym{2611} & \usym{2611}  & \usym{2611} &  \usym{2612} & \makecell{Superimposing\\ multiple images} \\
		
		\bottomrule[2pt]
	\end{tabular}
\end{table*}

\subsubsection{Face attribute}\

\textit{Synthesis.} Yang \textit{et al.} \cite{6247759} proposed a tensor-based 3D face geometry reconstruction method for modifying face expressions in videos. He \textit{et al.} \cite{8718508} proposed to use attribute classification constraints to guarantee the correct change of the required attributes, while introducing reconstruction learning to preserve attribute-excluding details to ensure that irrelevant attributes do not change. Wang \textit{et al.} \cite{Wang_2021_CVPR} ensured that the sensitive attributes of faces in images cannot be inferred by CV through inverting corresponding attributes or maximizing the uncertainty of the attributes, rather than removing them. Mirjalili \textit{et al.} proposed a new model for protecting multiple attribute privacy, such as gender, race, and age, while maintaining the matching performance of attribute features \cite{9201364}.\par

\textit{Adversarial perturbation.} Mirjalili \textit{et al.} \cite{8272743} proposed to use adversarial perturbation to enable gender attributes to be assessed by CV as flipped, i.e., females as judged as males and vice versa, and meanwhile, the face matcher is still available. On the other hand, the perturbation it produces is clearly visible to HV. Low \textit{et al.} \cite{9746848} focused on the privacy implications of face expression detection in video, proposing a framework seeking a universal perturbation that allows automatic micro-expression recognition to be confused, which is then injected into the video to protect the expressions in the video.\par

\subsubsection{Physiological signal}\

Chen \textit{et al.} \cite{7961722} proposed to eliminate the physiological signal from facial videos using Laplace pyramid representation so that the corresponding physiological signals can no longer be accurately measured, but the scheme may produce unrealistic videos in order to eliminate signals \cite{10.1145/3411764.3445719}. PulseEdit \cite{9680677} is proposed to protect rPPG signals in facial video by adversarial perturbation. It first generates a target rPPG signal, and then combines the target signal with the original one to solve an optimization problem and obtain the perturbation, which is added uniformly to the facial region. However, it does not run fast enough to support real-time processing, and is poorly defended against deep learning rPPG methods. Sun \textit{et al.} \cite{9806161} proposed to use a deep learning method to modify the rPPG signal in facial videos, achieving a faster and more effective protection method compared to PulseEdit.\par

\subsubsection{Other}\

Xu \textit{et al.} \cite{7350779} considered the identity privacy issue posed by the body and proposed to modify some body features, e.g., height, weight, and skin color, by putting in place while preserving the usability of body postures. Kuang \textit{et al.} \cite{10.1145/3503161.3548235} proposed to encode information such as the pose to protect the pedestrian image privacy, which can be visually strict to the original image, but looks natural.\par

One of the biggest threats of CV to visual content is the detection and extraction of biometric features, especially for faces. Therefore, the study of privacy protection schemes for biometrics has been the focus of the fight against CV adversaries and is able to be subdivided into more directions as described above compared to others. However, it is also possible that the source of privacy for visual content is non-biometric, such as text and content themselves.\par

\subsection{Solutions for Non-Biometric Protection}

\subsubsection{Scene text}\

Adversarial perturbation for non-sequential visual tasks (e.g., recognition \cite{10.1145/3343031.3351088} and detection \cite{9148656}) has been extensively studied in recent years. However, research on adversarial perturbation for sequential tasks (e.g., scene text recognition) is limited.\par

Yuan \textit{et al.} \cite{9162685} proposed an adversarial attack scheme based on multi-task learning to sequential tasks for preventing optical character recognition (OCR). Chen \textit{et al.} \cite{10.1007/978-3-030-67664-3_33} proposed to disguise the adversarial perturbation as a watermark to attack the OCR model. However, both schemes are for OCR, but optical characters are hard to see in everyday life, and often appear as text contained in natural images. Xu \textit{et al.} \cite{Xu_2020_CVPR} proposed an efficient and generalized adversarial method for scene text in the natural image, and attacked 5 state-of-art CV models, which are connectionist-based and attention-based temporal classification, to show the effectiveness of the method. However, the above methods are a white-box attack, which requires a complete understanding of the adversary's CV model, including (but not limited to) structure, parameters, and gradients, which is difficult in practice. Recently, a novel black-box perturbation generation method for scene text is proposed that requires only a prior knowledge of the adversary model output \cite{10045728}. \par

\subsubsection{Content}\

Liu \textit{et al.} \cite{liu2017protecting} proposed to apply adversarial perturbations to all contents in the image that can be detected by the CV model to make them invisible to the CV, while the scheme allows adjusting parameters to regulate the perturbation strength. However, the scheme indiscriminately applies permutations to all detectable content, reducing the usability of the image. Xue \textit{et al.} \cite{9148656} proposed a framework to first determine the sensitive information, then determine the location of the sensitive content in the image, and finally use adversarial perturbation to protect the privacy of the sensitive content. Duan \textit{et al.} \cite{Duan_2021_ICCV} proposed a novel perspective on adversarial perturbation, i.e., making perturbation by dropping information rather than adding, and experimental results showed successful blocking of CV detection to image content. Jiang \textit{et al.} \cite{10.1145/3343031.3351088} proposed a block-box adversarial perturbation method for video, which first generates perturbations experimentally by training the model on an image to reduce the number of queries to the video recognition model in order to increase efficiency. However, as mentioned above, most adversarial perturbation schemes, both white-box and block-box, require a large computational cost, which is not practical for the average user. Rajabi \textit{et al.} \cite{rajabi2021practicality} proposed a semantic adversarial perturbation scheme that blocks the detection of CV adversaries by superimposing multiple images together using cryptographic ideas, which is almost computationally costless and supports reversibility. \par

Existing non-biometric protection against CV adversaries is often based on adversarial perturbations, which ensures visual realism and imperceptibility. However, it is often computationally expensive and may be specific to a specific CV model and not valid for other models, i.e., it is not transferable. Although Rajabi \textit{et al.} \cite{rajabi2021practicality} proposed an extremely low-overhead and transferable adversarial perturbations, the results in terms of visual effects are poor. Future work may require further research on low overhead, transferability and high visual quality to further improve the feasibility of protection.

\begin{figure*}[htb]
	\setlength{\belowcaptionskip}{-0.5cm} 
	\centering 
	\includegraphics[scale=0.45]{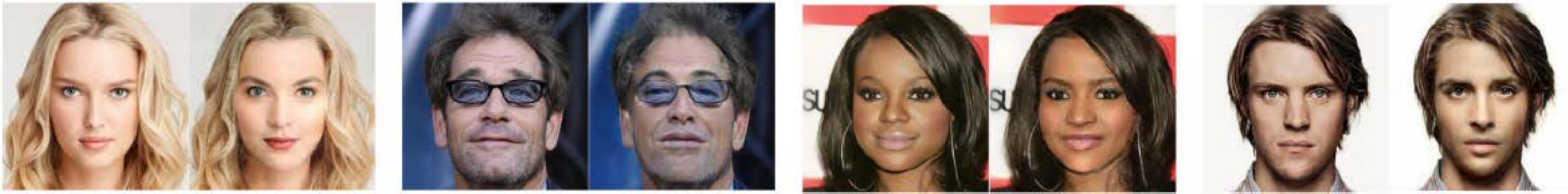} 
	\caption{Examples (from \cite{WEN2022197}) about the HV-invanriance is x, not equal to the ability to defend against HV adversaries. Left: the original image; eight: the image that identity is protected. For HV, the perceived content of the protected image does change somewhat compared to the original one, but not enough to affect HV's perception that it is the same identity.} 
	\label{HV-invariance}
\end{figure*}

\subsection{Common Principles}

Tables \ref{tab: cv-b} and \ref{tab: cv-nb} provide the breakdown of the reviewed solutions in biometric and non-biometric protection, respectively, in which HV-reality means that there is no obvious unnatural or false visual content, HV-invariance denotes that the information obtained from original visual content and protected one are the same for HV, and Type-I and V mean image and video, respectively. \usym{2611} represents that this item is full compliant with; \usym{2612} means that this item is not met; \usym{2610} implies that in some cases it may occur. It is should be noted that the fact that HV-invanriance is \usym{2612} does not mean that this is a defense against a HV adversary. An example is presented in Fig. \ref{HV-invariance}, in which HV does perceive that the protected image has changed, but it does not affect HV's ability to correctly perceive that the two identities are the same person.\par

By reviewing and summarizing the above solutions, 3 common principles can be identified.\par

\textbf{High precision.} Although CV adversaries can automate the execution of visual content understanding and extraction, they are often designed for specific tasks, i.e., they can only perform a single task using fixed features/contents associated with privacy in the visual content. Based on this, for the privacy protection tasks targeting CV, only the fixed features/contents need to be targeted. Other content unrelated to them can be maintained as is, thus achieving high-precision privacy protection.\par

\textbf{Transparent level.} It refers to the ability of visual content to be viewed naturally, i.e., without creating an unpleasant experience like blurring, while at the same preventing attacks on privacy by CV adversaries. This is because CV is still not comparable to HV, and its robustness is far worse than HV, and a small change (even imperceptible to the HV) can make CV fail.\par

\textbf{High usability.} Privacy protection schemes for CV adversaries tend to maintain high usability for CV and HV. This is a natural conclusion that can be drawn based on two principles mentioned above. For high precision, it ensures that non-privacy-related content does not change (or small change), and thus other content is often still understood and detected by CV. For the transparent level, it makes changes difficult to perceive for HV, ensuring that non-privacy content is correctly understood by HV, even for the content that is private to the CV adversary.\par

\begin{figure*}[htb]
	\setlength{\belowcaptionskip}{-0.5cm} 
	\centering 
	\includegraphics[scale=0.45]{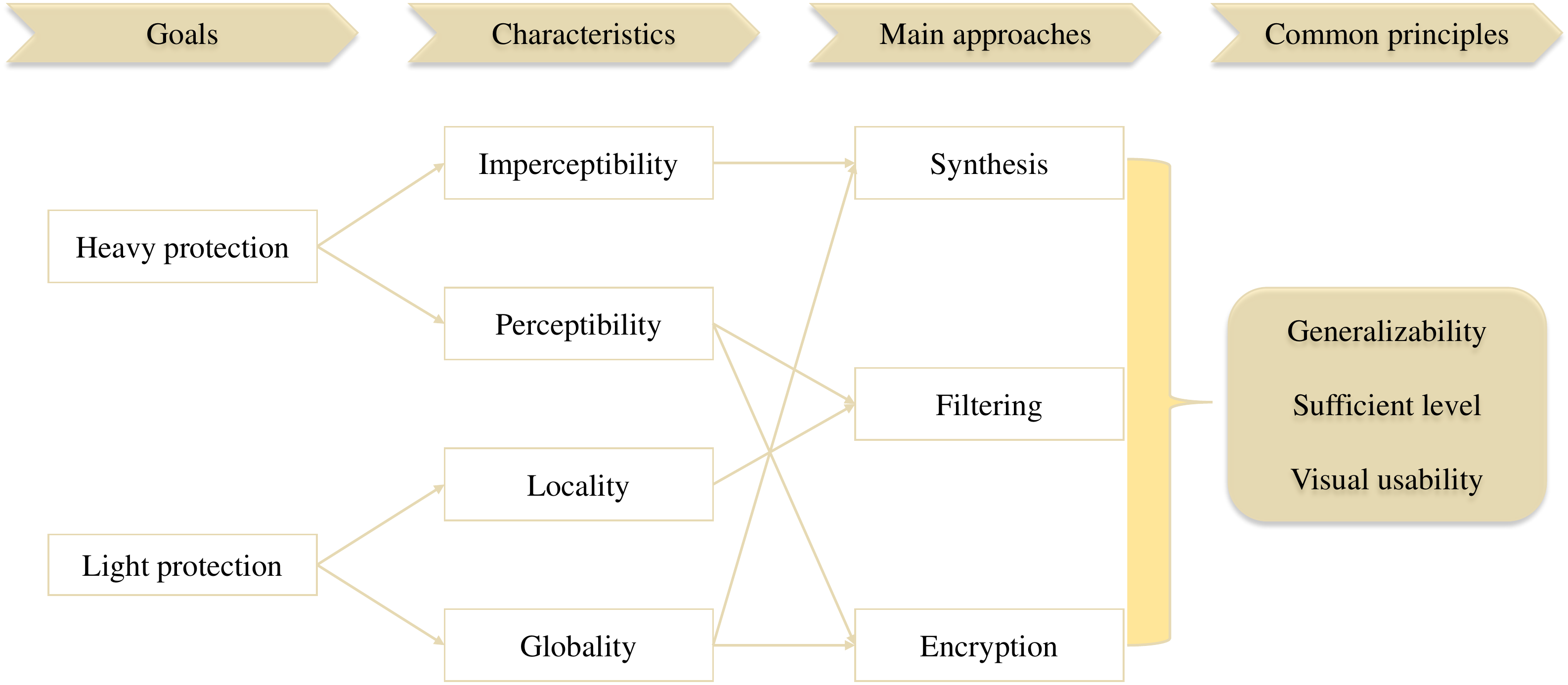} 
	\caption{Overview of the privacy protection analysis for HV adversary.} 
	\label{HV adversary framework}
\end{figure*}

\begin{figure*}[htb]
	\setlength{\belowcaptionskip}{-0.5cm} 
	\centering 
	\includegraphics[scale=0.45]{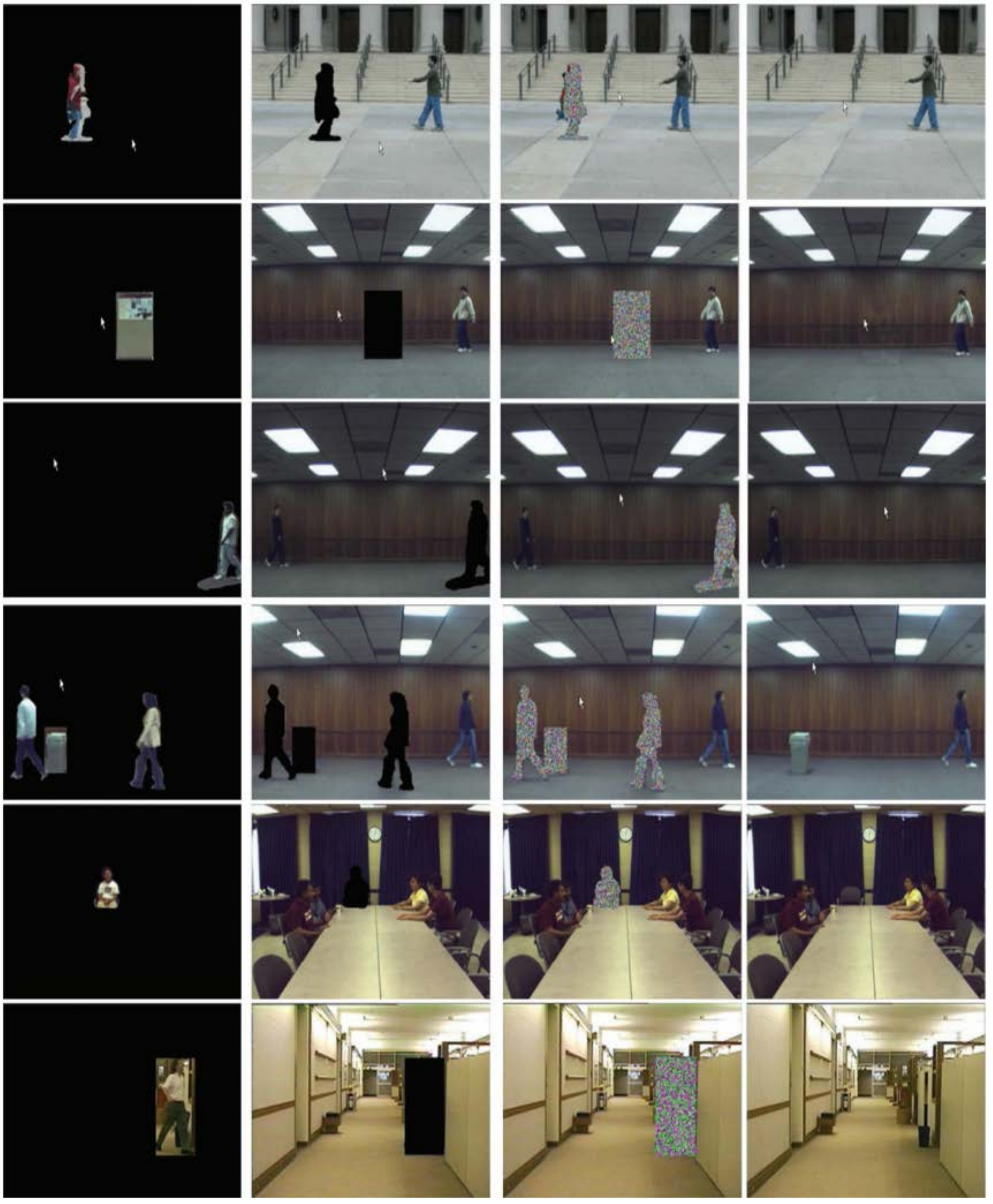} 
	\caption{Examples (from \cite{paruchuri2009video}) of heavy protection for HV adversaries. From left to right: privacy area, filtering, encryption, and synthesis.} 
	\label{heavy protection}
\end{figure*}

\begin{figure*}[htb]
	\setlength{\belowcaptionskip}{-0.5cm} 
	\centering 
	\includegraphics[scale=0.45]{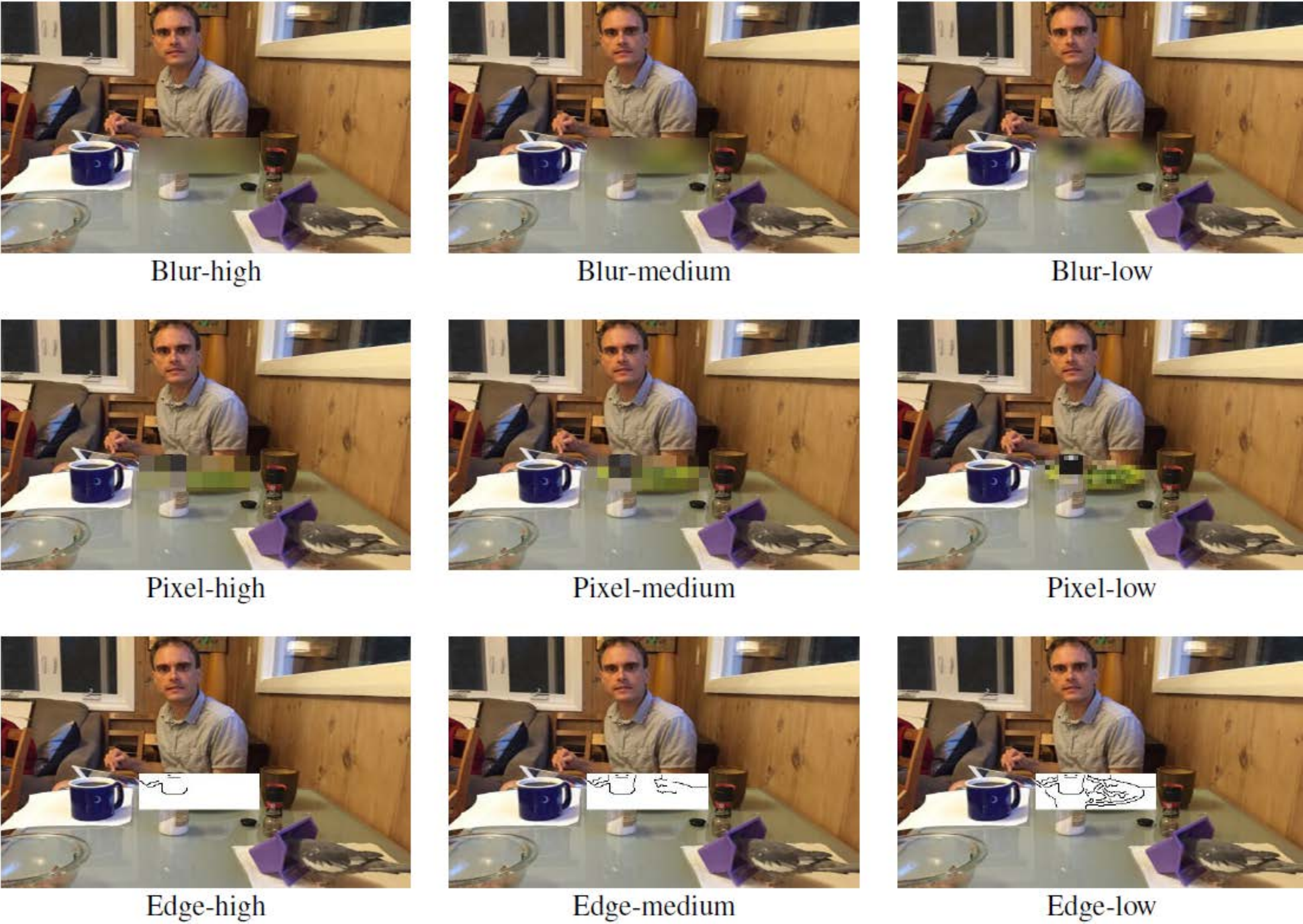} 
	\caption{Examples (from \cite{10.1145/3173574.3173621}) of light protection adjustable (which is local).} 
	\label{CV-light-example}
\end{figure*}

\begin{figure*}
	\centering 
	\subfigure[]{\includegraphics[scale=0.152]{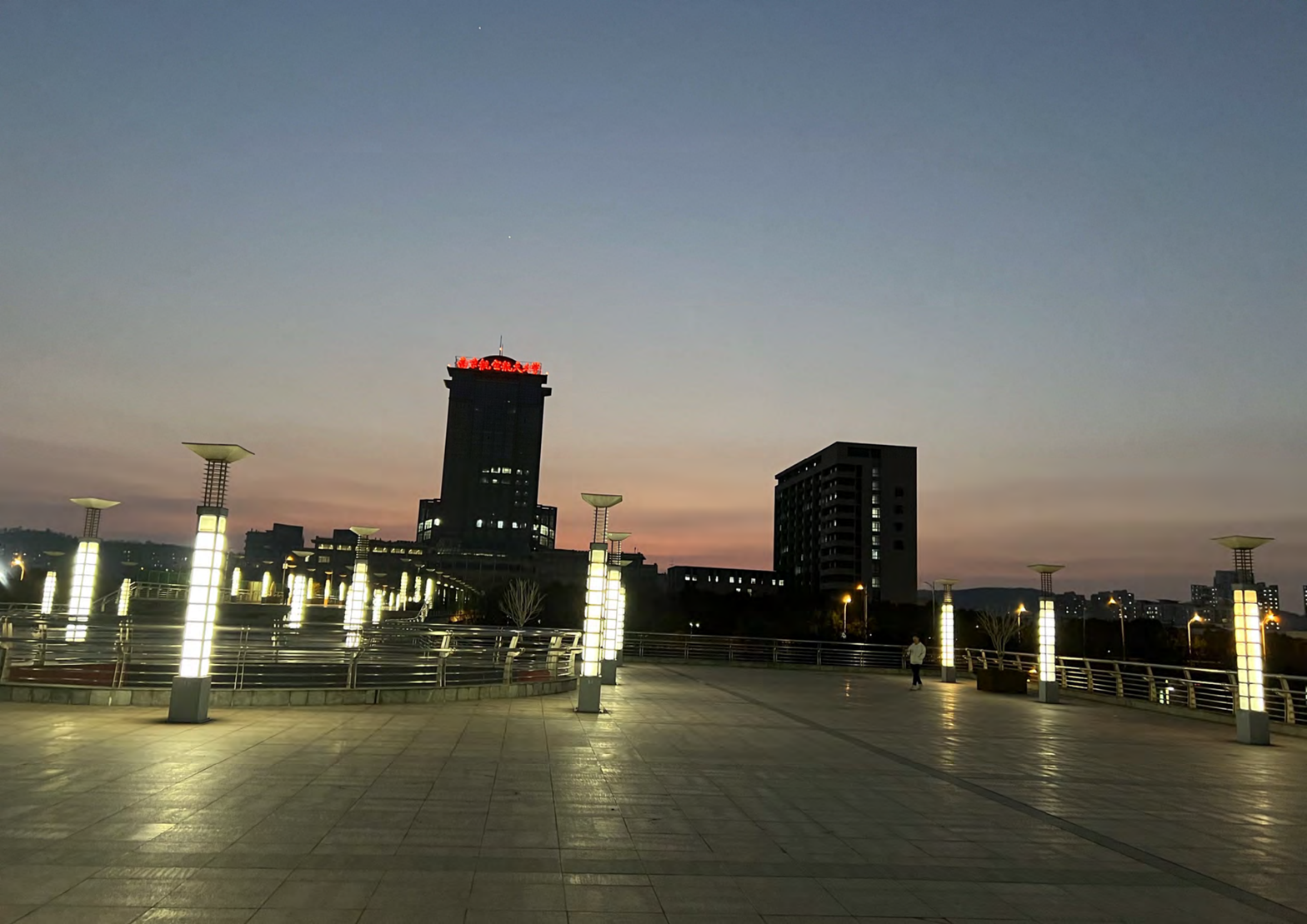}\label{original}}\vspace{-0.0em}
	\subfigure[]{\includegraphics[scale=0.03]{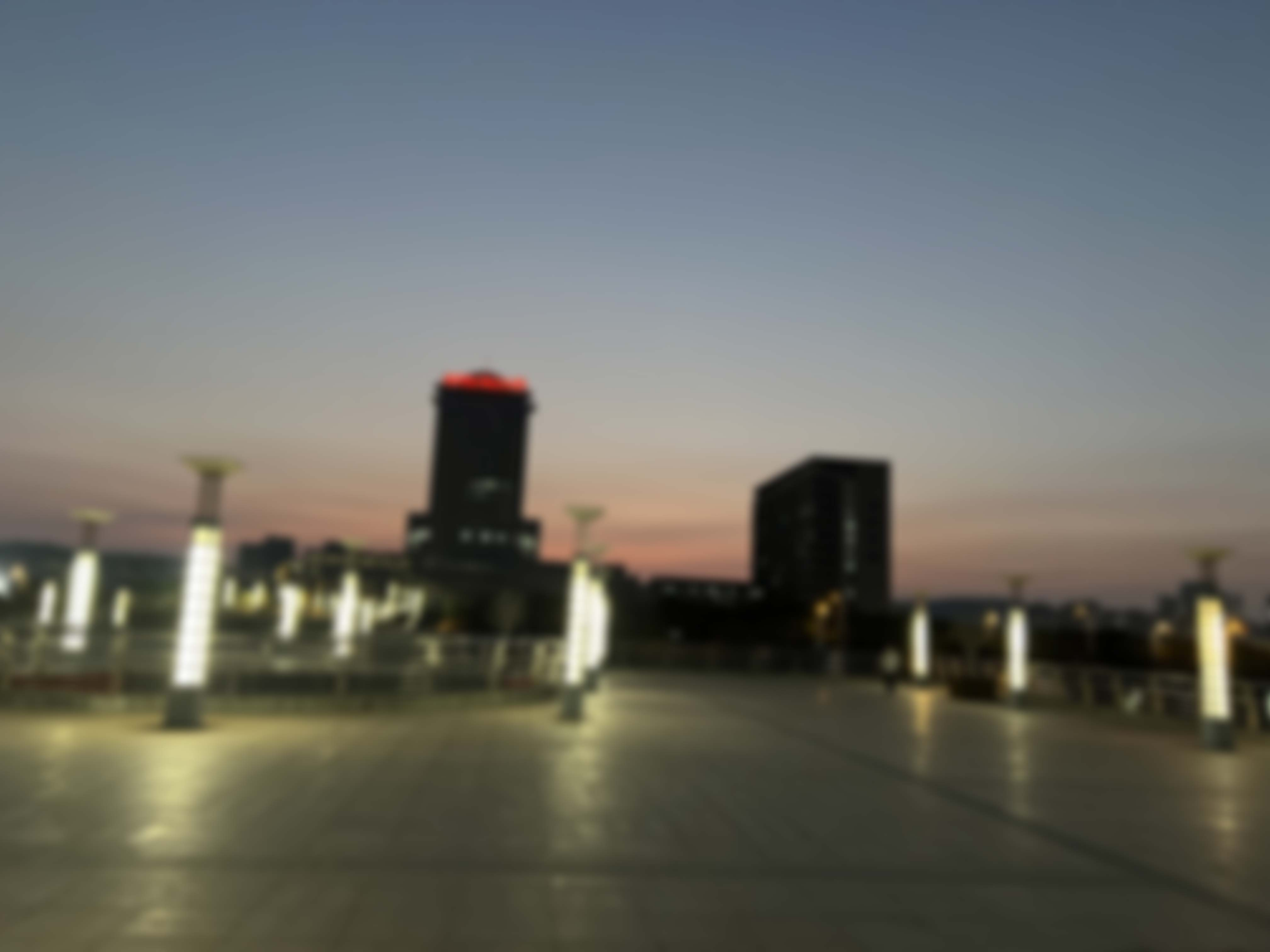}\label{blur}}
	\subfigure[]{\includegraphics[scale=0.03]{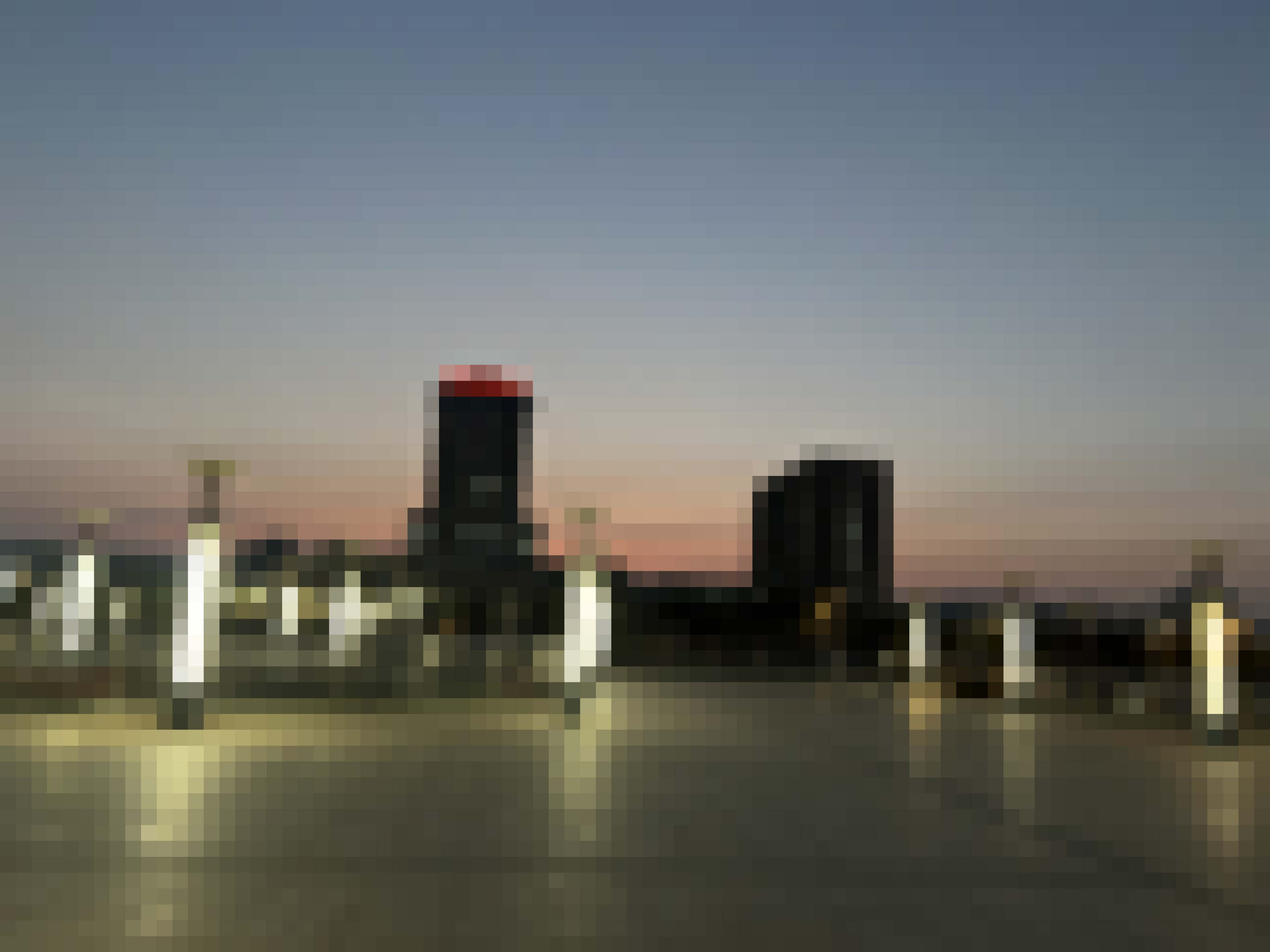}\label{pixlel}}
	
	\caption{Examples of typical filters. (a) original image; (b) blur; (c) pixelation.}
	\label{filer}
\end{figure*}

\section{Privacy Protection for HV Adversary} \label{Section: HV adversary}

This section provides a comprehensive analysis of privacy concerns posed by HV-based adversaries to visual content, including characteristics, main approaches, and common principles, as shown in Fig. \ref{HV adversary framework}. It should be noted that the HV is often very robust compared to CV perception, and it is difficult to change a feature or add noise to ht HV as in the case of CV adversary defense. In this part of the protection solution, it is often performed on objects or contents (which are collectively referred to areas), and thus is not classified according to privacy issues as in the case of CV adversary. \par

\subsection{Characteristics in HV Adversary Protection}

\subsubsection{Characteristics associated with heavy protection}
For heavy protection against the HV adversary, it means that the protected area does not reveal visual information. It is all local protection, as shown in Fig. \ref{heavy protection}, since if it is a global one, then it translates into confidentiality rather than privacy protection. Meanwhile, heavy protection is generally taken to two extremes, either HV clearly senses that a place has undergone severe abstraction (or distortion), or HV cannot detect it.

\begin{itemize}
	\item \textit{Imperceptibility.} It is a kind of visual privacy protection against HV adversaries that is imperceptible to the HV or difficult to detect. In general, this protection is refined and the content in the area is natural to HV and has semantic integrity (at the cost of misleading HV), as shown in the fourth column in Fig. \ref{heavy protection}.
	\item \textit{Perceptibility.} It refers to a very obvious perceptible treatment of a content or object, and although the HV knows that it is treated, no information can be extracted from the protected area, as shown in the second and third columns in Fig. \ref{heavy protection}. 
\end{itemize}

\subsubsection{Characteristics associated with light protection}

For light protection against the HV adversary, it means that the protected area still conveys the general properties of the original content, but the visual details have been erased. Also, how much of this general property is preserved in the protected content can often be adjusted by parameters as shown in Fig. \ref{CV-light-example}. This protection can be localized not only for an area as in heavy protection, but also for the visual content as a whole.

\begin{itemize}
	\item \textit{Locality.} It is a kind of processing containing rough content for the specific area of visual content, while the rest is maintained, which often occurs in every day, such as online chats and OSNs.
	\item \textit{Globality.} It mainly considers that the whole of the visual content is related to privacy, and therefore degrades all the content, which means that the protected content is a lower resolution version of the original one.
\end{itemize}

\subsection{Main Approaches}

\textit{Synthesis:} It has the same basic idea as `synthesis' in Section \ref{Section:3-2}, which is to replace privacy-sensitive content. However, the above section tends to be cautious about content changes is not sufficient to form a defense against HV. In this section, synthesis often apply object removal, replacement with abstract content, and regeneration that changes the semantic-level understanding.\par

\textit{Filtering:} It is one of the most common methods of visual content privacy protection in daily life due to its simplicity, efficiency, and effectiveness. As shown in Fig. \ref{filer}, two common filters, blur and pixelation, effects are demonstrated. Blurring is the application of a Gaussian function to the visual content, which uses adjacent pixels to modify each pixel. Pixelation is the process of dividing the visual content into square grids, calculating the average color of the pixels inside each grid, and assigning the color to all pixels in that grid.\par

\textit{Encryption:} In cryptography, encryption is the encoding of data in such a way that the original data becomes incomprehensible. Classical image and video encryption also aim to completely eliminate the comprehensibility of the visual content, that is, to guarantee confidentiality. For privacy protection, the complete elimination of comprehensibility is not necessary. It is possible to preserve some comprehensibility of visual content by selecting to encrypt some areas or by visual encryption.\par

\subsection{Solutions for Heavy Protection}

\subsubsection{Imperceptibility}\

Object (or people) removal is aims to protect privacy by hiding objects so that no trace of them appears in the protected visual content \cite{elharrouss2020image}. After an object is removed, a gap appears, which needs to be inpainted. Inpainting means reconstructing the missing part using an imperceptible way, i.e., using information from the surrounding area to fill in the missing part.\par

Barnes \textit{et al.} \cite{10.1145/1531326.1531330} proposed to search for similar patches in known regions to synthesize the missing content parts step by step, where the authors designed an approximate nearest-neighbor search to find the closest patches to match. Although it is good at filling high-frequency textures, it cannot get the support of the overall structure of the visual content. Recently, deep convolutional networks (CNNs) have shown great potential for visual content inpainting \cite{elharrouss2020image}. CNN-based inpainting techniques can predict the missing parts based on the surrounding environment, and thanks to the increase in computing power, large-scale data training is possible, making it possible to produce semantically sound results. However, the result may lack surprising texture details and be blurry. Yan \textit{et al.} \cite{Yan_2018_ECCV} combined the advantages of patch-based and CNN-based methods, and proposed to apply U-Net \cite{10.1007/978-3-319-24574-4_28} as the backbone network and add the shift-connection layer to U-Net. Then, the features of known parts are used to shift to the missing parts, thus generating semantically sound and finely textured inpainting results.\par

Uittenbogaard \textit{et al.} \cite{Uittenbogaard_2019_CVPR} are concerned about the millions of images hosted in Street View maps, which contain sensitive objects such faces and license plates. They suggested that the use of methods such as blurring might not allow for adequate protection of privacy, and argued that sensitive objects should be removed. They proposed a privacy-preserving scheme that automatically segments and removes the moving objects in street view maps, and impaints the missing parts. Nakamura \textit{et al.} \cite{8270072} proposed that the text appearing in images may carry private information, and to protect privacy by erasing them. They suggested that instead of extracting the text precisely, only need to know the approximate location of the text, and then apply the CNN model to fill the text area with the color of the surrounding pixels. However, this scheme is an indiscriminate removal of all text from the image and is not flexible. Tursun \textit{et al.} \cite{8978083} proposed a conditional GAN with an auxiliary mask, capable of removing all or user-specified text. \par

More recently, for the protection of scene text against HV adversaries, researchers have proposed to replace text. Roy \textit{et al.} \cite{Roy_2020_CVPR} proposed a font adaptive network for modifying individual characters in the scene text directly on the image, achieving the same structure as the source font and preserving the original color. Then, a work to modify the scene text on a video is proposed, which works by selecting a frame as a reference in which the text is modified, and then transferring the text to other frames using a newly designed network \cite{G_2021_ICCV}. Yu \textit{et al.} \cite{s21010058} proposed to apply differential privacy combined with GAN to protect privacy in IoT, where the protection for license plate numbers is mentioned. That is, the original license plate number is extracted as a feature, Laplace noise is added to it, and the rule-compliant license plate number is regenerated.\par

Imperceptible privacy protection is difficult for adversaries to distinguish between what visual content is protected and what is not. Meanwhile, it also ensures the comfortable viewing of visual content. However, this is a double-edged sword for users, making them equally unaware of whether the content has changed and whether their understanding of the content is correct.

\begin{figure*}[htb]
	\setlength{\belowcaptionskip}{-0.5cm} 
	\centering 
	\includegraphics[scale=0.5]{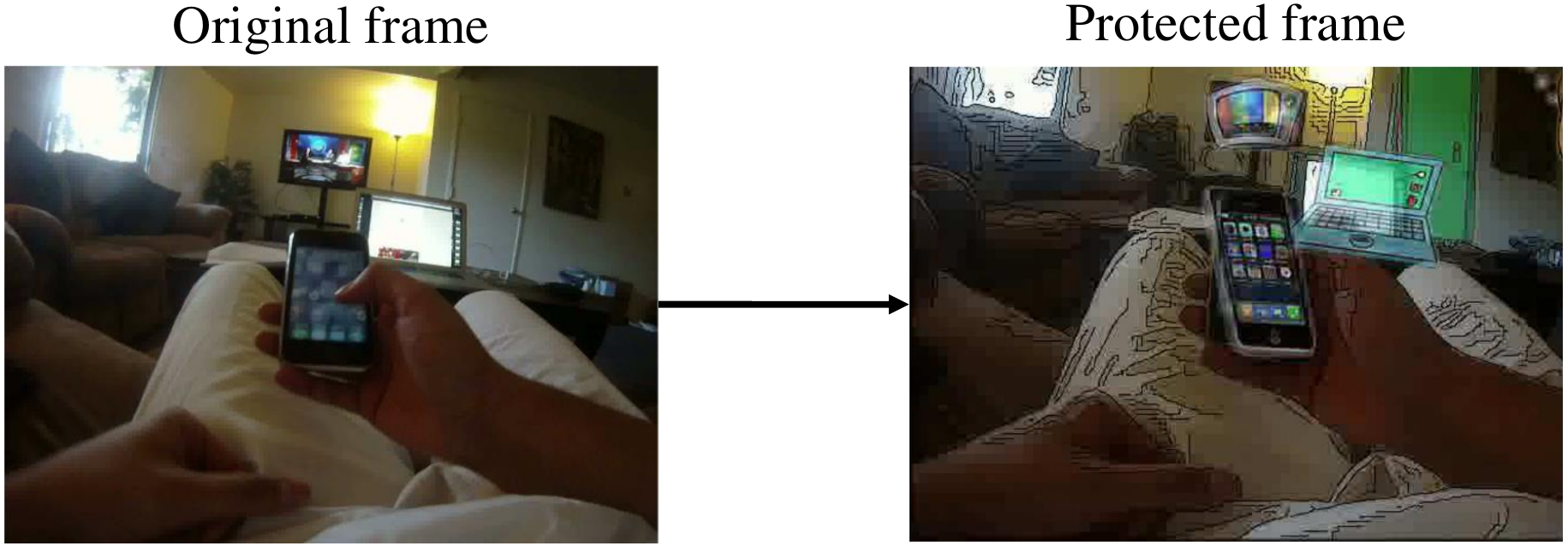} 
	\caption{An example (from \cite{Kapadia_2017_CVPR_Workshops}) of cartooning to protect privacy.} 
	\label{cartoon}
\end{figure*}

\begin{figure*}[htb]
	\setlength{\belowcaptionskip}{-0.5cm} 
	\centering 
	\includegraphics[scale=0.4]{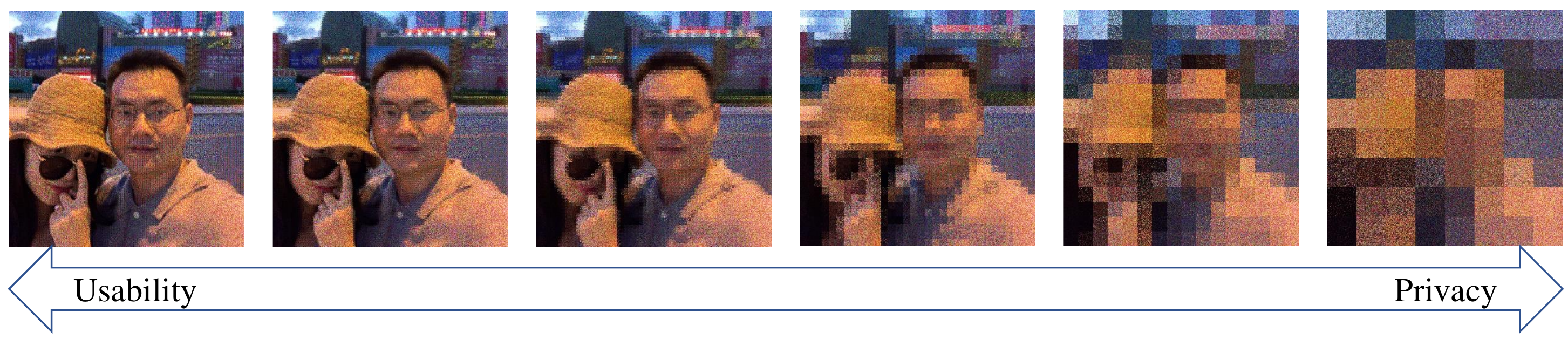} 
	\caption{An example (from \cite{9393403}) of tunable balance between privacy and visual usability in TPE schemes.} 
	\label{TPE-diff}
\end{figure*}

\subsubsection{Perceptibility}\

\textit{Filtering.} Chinomi \textit{et al.} \cite{10.1007/978-3-540-77409-9_14} proposed a privacy scheme using visual abstraction to protect specific objects in video surveillance, where the relationship between viewer and object is considered. Different viewers will view different levels of visual abstraction, e.g., silhouette and edge, which is a kind of adaptive privacy protection. Orekondy \textit{et al.} \cite{Orekondy_2018_CVPR} proposed an image redaction scheme for privacy protection. It is able to label multiple privacy regions, which can be considered as an auto-tagging task for multiple privacy attributes, and segment them automatically. Then, these regions are completely masked using a heavy filter (like the second column in Fig. \ref{heavy protection}). Meanwhile, they are then labeled with text in order to make it easier for the user to know what is in them to prevent misinterpretation. For example, the face area is masked and labeled with `face'. Bonetto \textit{et al.} \cite{7285023} used filters for surveillance videos captured by small UAVs to process the privacy regions in them, where masking filters were used.\par

\textit{Encryption.} Concerned about the invasion of human privacy by surveillance video,  Boult \cite{1623767} proposed an encryption scheme for the human face area, which allows surveillance of a general nature while improving privacy issues, and full access by key in the event of an accident. Dufaux \textit{et al.} \cite{4559592} considered the format compatibility, and proposed a region-of-interest (ROI) video encryption scheme, based on transform-domain or codestream-domain, suitable for MPEG-4. Users expect different levels of privacy for viewing the images they share in OSNs for different people. Aribilola \textit{et al.} \cite{9983830} considered the privacy impact of video captured by dynamic surveillance cameras and proposed a low computationally intensive unsupervised learning for accurate motion detection in dynamic surveillance video. Then, the detected sensitive content (i.e., ROI) is encrypted using lightweight Chacha 20 cipher that is specifically used in IoT devices. In addition, He \textit{et al.} \cite{7579755} proposed to be able to encrypt one or more ROIs of an image while distributing different keys to different viewers, enabling them to see the image with different decryption levels. \par

Perceptible protection ensures that HV adversaries are completely unaware of the content of sensitive areas, while ensuring that users know that the areas are protected. However, the effect of heavy protection is very obvious, and viewing is often very unattractive. Meanwhile, the sensitive areas do not reveal any visual information to the user, which sometimes can be a nuisance.\par

\subsection{Solutions for Light Protection}

\subsubsection{Locality}\

\textit{Filtering.} The large amount of realistic images collected in Google Street View has the huge side effect of potentially carrying private information such as faces and license plates. Therefore, Google applies the blurring filter method to blur the privacy area detected therein to protect privacy \cite{5459413}. Brkic \textit{et al.} \cite{8014907} considered the privacy problem due to the presence of the human body in the image and proposed a scheme to blur the face and body, which not only protects biometric but also non-biometric features. Irrelevant faces in live streaming often inevitably appear in the footage, damaging their privacy. Zhou \textit{et al.} \cite{9218980} proposed a pixelated protection scheme for irrelevant faces in live streaming. The scheme first detects faces on a single frame and generates the corresponding face vectors, then quickly associates the same faces between frames to generate face track vectors and refine them, and finally pixelates the faces. Sun \textit{et al.} \cite{sun2022zoomp} considered the privacy issues associated with videoconferencing, such as videoconferencing recordings to record a face. They proposed the use of the blurring filter to protect privacy.\par

\subsubsection{Globality}\

\textit{Synthesis.} Erdélyi \textit{et al.} \cite{6918642} considered images (e.g., captured by surveillance cameras) as a whole to be private and proposed transforming the original image into an abstract cartoon in order to erase the sensitive details, while this scheme allows the intensity of protection to be changed according to different parameters. Note that the authors claimed to be able to use it in videos, but it is actually only done on one frame. Subsequently, the authors \cite{erdelyi2014multi} further considered that different contents in an image have different privacy levels and proposed that the protection intensity is greater for highly privacy-sensitive contents and becomes lower for the rest. Hassan \textit{et al.} \cite{Kapadia_2017_CVPR_Workshops} proposed a content replacement scheme on video streams to protect privacy, i.e., using cartoonish objects and backgrounds (which erase the details of the original content and are an abstract representation) to replace the original contents. As shown in Fig. \ref{cartoon}, everything in the original visual content is transformed into cartoon form, erasing the real details.\par

\textit{Encryption.} Traditional visual content encryption often transforms an image or video into a form without any visual meaning, making it difficult to view and lacking in visual usability. Recently, some new types of encryption with visual effects have been proposed. Bao \textit{et al.} \cite{BAO2015197} proposed the concept of visually meaningful image encryption, that is, encrypting an image using a traditional scheme and then hiding it within a carrier image using data hiding. As a result, the original image is protected, but the final result is visually appealing. However, visual effects have no relationship to the original image, leaving visual usability still missing. To address this, Zhao \textit{et al.} \cite{ZHAO2022628} proposed the primitively visually meaningful image encryption, which means that the encrypted image carries the rough visual information of the original image, without revealing the fine content, to ensure visual usability. Wright \textit{et al.} \cite{10.1145/2756601.2756618} proposed the idea of thumbnail-preserving encryption (TPE), which means that the thumbnail of the encrypted image is the same or nearly the same as the original thumbnail. Meanwhile, the information preserved in the encrypted image can be easily adjusted according to the parameters as shown in Fig. \ref{TPE-diff}, thus balancing privacy with visual usability. Then, they designed the first TPE scheme by utilizing the permutation-only encryption. However, some cryptographic studies \cite{7295616} show that permutation-only encryption is not unbreakable. Tajik \textit{et al.} \cite{2019NDSS} proposed the first TPE scheme with theoretical security support, which applies the substitution-permutation encryption framework commonly used in traditional cryptography, and proved the corresponding security. Based on this, Zhang \textit{et al.} \cite{9393403} proposed to utilize the idea of data hiding to construct a TPE scheme with security guarantee.\par

Light protection is relatively better visually and has a better browsing experience compared to heavy protection. However, it also has the obvious side effect that adversaries may also infer the content from the rough information preserved in the protected visual content.\par

\begin{table*}[]
	\centering
	\caption{A summary of the solutions in heavy protection.}
	\renewcommand{\arraystretch}{1.5}
	\tabcolsep 7pt
	\label{tab: hv-hp}
	\begin{tabular}{ccccccccc}
		\toprule[2pt]
		\multirow{2}{*}{\makecell{Sub\\ class}} & \multirow{2}{*}{Paper} & \multirow{2}{*}{Year} & \multirow{2}{*}{\makecell{Key \\technology}} & \multirow{2}{*}{Adjustability} & \multirow{2}{*}{Reversibility} & \multicolumn{2}{l}{Type} & \multirow{2}{*}{Note} \\ \cline{7-8}
		&                        &                       &                      &                  &              & I           & V          &                       \\ \hline
		
		\multirow{8}{*}{\makecell{Impercep-\\tibility}}  & \cite{10.1145/1531326.1531330} & 2009 & Inpainting & \usym{2612} & \usym{2612} & \usym{2611} & \usym{2612} & Based on patches \\
		& \cite{Yan_2018_ECCV} & 2018 & Inpainting & \usym{2612} & \usym{2612} & \usym{2611} & \usym{2612} & Based on CNNs \\
		& \cite{Uittenbogaard_2019_CVPR} & 2019 & \makecell{Object\\ removal} & \usym{2612} & \usym{2612} & \usym{2611} & \usym{2612} & \makecell{Focused on\\ view street} \\
		& \cite{8270072} & 2017 & Text removal & \usym{2612} & \usym{2612} & \usym{2611} & \usym{2612} & Based on CNNs \\
		& \cite{8978083} & 2019 & Text removal & \usym{2611} & \usym{2612} & \usym{2611} & \usym{2612} & \makecell{Focused on\\ user-specified text} \\
		& \cite{Roy_2020_CVPR} & 2020 & \makecell{Text\\ modification} & \usym{2611}  & \usym{2612} & \usym{2611}  & \usym{2612} &  \makecell{Characteristic-level\\ modification} \\
		& \cite{G_2021_ICCV} & 2021 & \makecell{Text\\ modification} & \usym{2611} & \usym{2612} & \usym{2612} & \usym{2611} & \makecell{Frame-level\\ modification} \\
		& \cite{s21010058} & 2021 & GAN & \usym{2612} & \usym{2612} & \usym{2611} &  \usym{2612} & \makecell{Differential\\ privacy} \\
		\hline
		\multirow{7}{*}{\makecell{Percep-\\tibility}}          &    \cite{10.1007/978-3-540-77409-9_14} &  2008 & \makecell{Visual\\ abstraction} & \usym{2611} & \usym{2612} & \usym{2612} & \usym{2611} & \makecell{Different effects\\ for different users}\\
		& \cite{Orekondy_2018_CVPR} & 2018 & Redaction & \usym{2612} &  \usym{2612} & \usym{2611} & \usym{2612} & \makecell{Multiple privacy\\ attributes protection} \\
		& \cite{7285023} & 2015 & \makecell{Visual\\ abstraction} &\usym{2611} & \usym{2612} & \usym{2612} & \usym{2611} & \makecell{Focused on\\ drone video} \\
		& \cite{1623767} & 2005 & Encryption & \usym{2612} & \usym{2611} & \usym{2612} & \usym{2611} & \makecell{Focused on\\ face} \\
		&  \cite{4559592}  & 2008 & Encryption & \usym{2612} & \usym{2611} & \usym{2612} & \usym{2611} &  \makecell{Compatible with\\ MPEG-4} \\
		& \cite{9983830} & 2022 & Encryption &  \usym{2612} & \usym{2611} & \usym{2612} & \usym{2611} & \makecell{Based on\\ Chacha20 cipher} \\
		& \cite{7579755} & 2016 & Encryption & \usym{2611} & \usym{2611} & \usym{2611} & \usym{2612} & \makecell{Different effects\\ for different viewers} \\
		\bottomrule[2pt]
	\end{tabular}
\end{table*}

\begin{table*}[htb]
	\centering
	\caption{A summary of the solutions in light protection.}
	\renewcommand{\arraystretch}{1.5}
	\tabcolsep 7pt
	\label{tab: hv-lp}
	\begin{tabular}{ccccccccc}
		\toprule[2pt]
		\multirow{2}{*}{\makecell{Sub\\ class}} & \multirow{2}{*}{Paper} & \multirow{2}{*}{Year} & \multirow{2}{*}{\makecell{Key \\technology}} & \multirow{2}{*}{Adjustability} & \multirow{2}{*}{Reversibility} & \multicolumn{2}{l}{Type} & \multirow{2}{*}{Note} \\ \cline{7-8}
		&                        &                       &                      &                  &              & I           & V          &                       \\ \hline
		
		\multirow{4}{*}{\makecell{Locality}}  & \cite{5459413} & 2009 & Blur & \usym{2612} & \usym{2612} & \usym{2611} & \usym{2612} & \makecell{Focused on\\ view street} \\
		& \cite{8014907} & 2017 & Blur & \usym{2611} & \usym{2612} & \usym{2611} & \usym{2612} & \makecell{Focused on\\ body and face} \\
		& \cite{9218980} & 2021 & Pixelation & \usym{2612} & \usym{2612} & \usym{2612} & \usym{2611} & \makecell{Focused on\\ live streaming} \\
		& \cite{sun2022zoomp} & 2022 & Blur & \usym{2612} & \usym{2612} & \usym{2612} & \usym{2611} & \makecell{Focused on\\ video conference} \\
		\hline
		\multirow{8}{*}{\makecell{Globality}}          &    \cite{6918642} &  2014 & \makecell{Cartoonli-\\zation} & \usym{2611} & \usym{2612} & \usym{2611} & \usym{2612} & \makecell{Homogeneous\\ protection}\\
		&    \cite{erdelyi2014multi} &  2014 & \makecell{Cartoonli-\\zation} & \usym{2611} & \usym{2612} & \usym{2611} & \usym{2612} & \makecell{Differentiated\\ protection}\\
		& \cite{Kapadia_2017_CVPR_Workshops} & 2017 & \makecell{Cartoonli-\\zation} & \usym{2611} &  \usym{2612} & \usym{2612} & \usym{2611} & \makecell{Objection\\  replacement} \\

		& \cite{BAO2015197} & 2015 & Encryption &\usym{2612} & \usym{2611} & \usym{2611} & \usym{2612} & \makecell{Visual effect\\ irrelevant} \\
		& \cite{ZHAO2022628} & 2022 & Encryption & \usym{2611} & \usym{2611} & \usym{2611} & \usym{2612} & \makecell{Visual effect\\ relevant} \\
		&  \cite{10.1145/2756601.2756618}  & 2015 & Encryption & \usym{2611} & \usym{2611} & \usym{2611} & \usym{2612} &  Insecurity \\
		& \cite{2019NDSS} & 2019 & Encryption &  \usym{2611} & \usym{2611} & \usym{2611} & \usym{2612} & \makecell{Security\\ guarantee} \\
		& \cite{9393403} & 2022 & Encryption & \usym{2611} & \usym{2611} & \usym{2611} & \usym{2612} & \makecell{Based on\\ data hiding} \\
		\bottomrule[2pt]
	\end{tabular}
\end{table*}

\subsection{Common Principles}

Tables \ref{tab: hv-hp} and \ref{tab: hv-lp} provide the breakdown of the reviewed solutions in heavy and light protection, respectively, in which adjustability means that the extent of the protected area, the content shown can change according to different parameters, or different people can have differentiated visual effect.\par

By reviewing and summarizing the above solutions, 3 common principles can be identified.\par

\textbf{Generalization.} In contrast to CV adversaries, protection against HV adversaries tends to specify an area to modify, rather than first extracting features from the visual content, modifying the features, and then regenerating according to the features. This makes protection schemes often have generalized characteristics, making that they protect all the details of an entire area, rather than being finely targeted.\par

\textbf{Sufficient level.} Sufficient protection of visible information about sensitive content is required since HV is far more complex and robust than CV. That is, fine information about sensitive content requires to be erased, and even rough information cannot be preserved if necessary. In general, there are two very different ways to implement this principle: one is to erase sensitive content and generate realistic fake content; the other is to use a visually distinctive method, such as a filter, to obscure sensitive content.\par

\textbf{Visual usability.} Protection against HV adversaries, whether heavy or light, it preserves a degree of visual usability that otherwise translates into a form of confidentiality protection. For heavy protection, it is all a localized protection, which means that the visual content of the unprotected area can be used normally. For light protection, the protected private content itself comes with some visual usability, but it is the fine content is erased.\par

\begin{figure*}[htb]
	\setlength{\belowcaptionskip}{-0.5cm} 
	\centering 
	\includegraphics[scale=0.45]{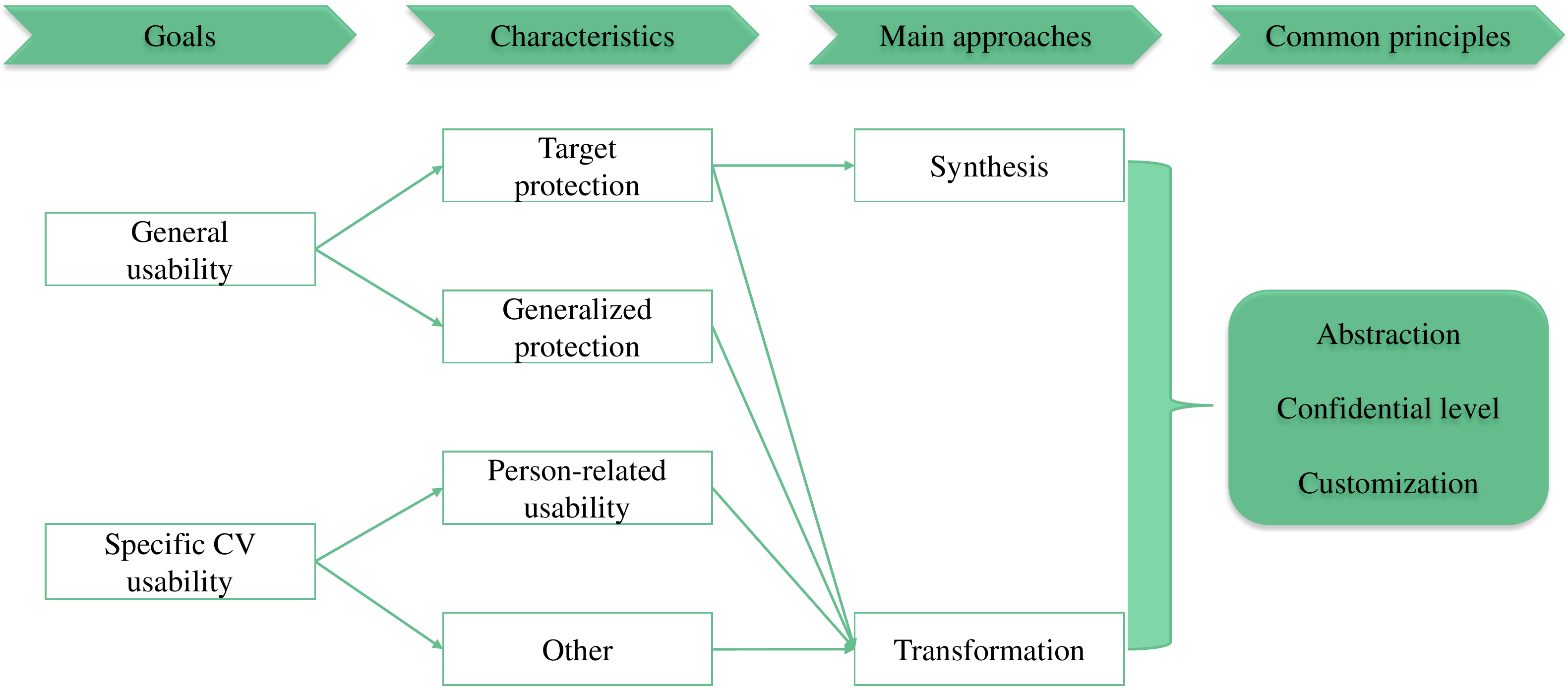} 
	\caption{Overview of the privacy protection analysis for CV\&HV adversary.} 
	\label{mixed adversary framework}
\end{figure*}

\section{Privacy Protection for CV\&HV Adversary} \label{Section: mixed adversary}

This section provides a comprehensive analysis of privacy concerns posed by CV\&HV-based adversaries to visual content, including characteristics, main approaches, and common principles, as shown in Fig. \ref{mixed adversary framework}. Compared to the above two adversaries, this section has relatively little work.\par

\subsection{Characteristics in HV\&CV Adversary Protection}

\subsubsection{Characteristics associated with general usability}

For general usability protection, it means that the protected visual content also preserves a certain visual effect to ensure the HV usability and that is not weaker than the CV usability. Meanwhile, CV models may be able to perform specific tasks, e.g., detection, on protected visual content.\par

\begin{itemize}
	\item \textit{Target protection.} It refers to finding the privacy area in the visual content to be protected, processing it to remove private information, while preserving certain features to ensure that the CV can perform a specific task.
	
	\item \textit{Generalized protection.} It refers to the processing of the visual content as a whole, rather than some specific areas, to remove the maximum amount of private information. It tends to ensure that neither the HV nor CV can access or even reason about the private information, while still supporting other effective analyses and perception.
\end{itemize}

\subsubsection{Characteristics associated with specific CV usability}\

For Specific CV usability protection, it implies that the visual content contains almost no information usable to HV, only usable for a limited CV task or a specific CV model, while this usability often requires prior adversarial training of CV using protected data. This part considers the dilemma of wanting a CV to be able to perform an important task while preventing it from gaining unnecessary information that could lead to a privacy breach. \par

\begin{itemize}
	\item \textit{Person-related usability.} It considers the usability associated with people in visual content, in which CV performs recognition of human body behavior, faces, etc. The visual content in this sub-part is often considered  surveillance video, which needless to say also contains a lot of private information, and also requires that they also have the need to analyze the necessary information about in it, such as action recognition.
	
	\item \textit{Other.} It is protects not only the privacy in a way, but also the security of the data itself, preventing the data from being used by unauthorized CV models or from being to perform unintended tasks.
\end{itemize}

\subsection{Main Approaches}

\textit{Synthesis:} Compared to the `synthesis' in the two sections above, the synthesis in this section is more radical, like an abstraction, i.e., all unnecessary information is removed, e.g., color, texture, and shape, and only the information necessary to perform a certain task is preserved in the re-generated content. It is important to note that the necessary information preserved is not only usable to CV, but can also be used by HV.\par

\textit{Transformation:} It is meant to convert data from one format to another one. In this part, the original visual content is often converted into a different style to erase all real details or directly into a form that is not perceptible to HV while preserving the necessary CV usable information.\par

\subsection{Solutions for General Usability}

\subsubsection{Target protection}\

\textit{Synthesis.} Zou \textit{et al.} \cite{9747456} considered the importance of  pedestrian action in videos, but also carry unnecessary information such as identity. They proposed to protect their privacy while supporting action recognition by replacing human bodies with their pose models. Kunchala \textit{et al.} \cite{Kunchala_2023_WACV} likewise noted that the large amount of pedestrian privacy contained in the video and proposed using avatars to replace real bodies. The avatar was synthesized using a 3D body while integrated with digital inpainting, erasing all personal information while retaining each person's presence, pose, shape, and background information. Yang \textit{et al.} \cite{yang2022digital} proposed the use of digital masks to ensure patient face privacy while preserving the necessary information needed for diagnosis. Experiments in disease diagnosis have shown comparable accuracy using digital masks and raw images.\par

\textit{Transformation.} Brkić \textit{et al.} \cite{BRKIC201741} proposed a scheme for de-identifying humans in surveillance videos, while supporting CV detection and segmentation to humans. Specifically, human locations are first detected using background subtraction based on Gaussian mixture modes, and then rendered using a deep neural network to remove biometric and non-biometric features about the person (including face, clothes, hair, etc.). This makes face recognition and identity detection no longer feasible, but retains the role of overall human body detection and segmentation.

\begin{figure*}[htb]
	\setlength{\belowcaptionskip}{-0.5cm} 
	\centering 
	\includegraphics[scale=0.5]{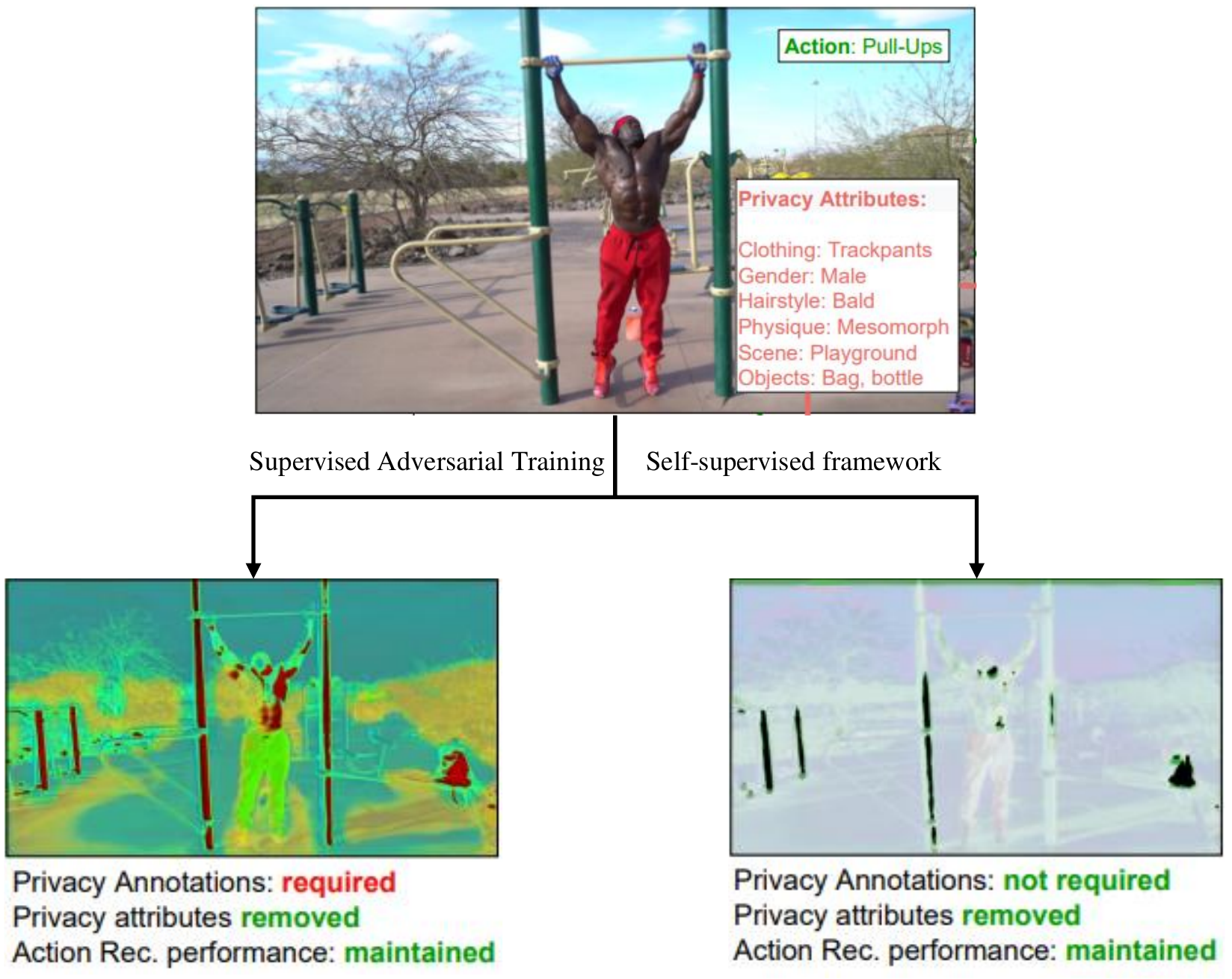} 
	\caption{Overview (from \cite{Dave_2022_CVPR}) of the privacy-preserving action recognition scheme \cite{9207852} (left) and \cite{Dave_2022_CVPR} (right). Both schemes enable action recognition, but \cite{9207852} requires labels and \cite{Dave_2022_CVPR} does not.} 
	\label{action recognition}
\end{figure*}

Targeted protection is suitable for scenarios where a clear protection target is known, and where best efforts are made to remove the privacy of the target while preserving its specific usability. However, in real-world scenarios, privacy is often unexpected, meaning that it is difficult to say that there is no privacy threat to something outside a specific target, and targeted protection to deal with such threats is difficult. 

\subsubsection{Generalized protection}\

Ryoo \textit{et al.} \cite{Ryoo_Rothrock_Fleming_Yang_2017} proposed that extreme low-resolution can be used to anonymize video and prevent CV and HV to recognize the identity of the person appeared in the video. Meanwhile, the paradigm of inverse super resolution was proposed to ensure the usability of video for activity recognition. Specifically, the aim of the paradigm is to perform a learning transformation from a set of high-resolution images such that the low-resolution images which is generated maintain a comparable amount of information as the high-resolution ones. The experimentation verified that the scheme is successful in activity recognition. Hinojosa \textit{et al.} \cite{Hinojosa_2021_ICCV} considered the direct design of privacy-preserving cameras that capture only useful information to sense people in the scene while hiding other privacy-sensitive information, making the captured image also an extreme low resolution.\par

Wu \textit{et al.} \cite{9207852} proposed a novel framework of the adversarial training to balance privacy and usability of video content, achieving privacy-preserving action recognition. However, this adversarial training requires training data containing privacy labels, which requires a huge cost to annotate the raw data. Meanwhile, the scheme based on privacy label training are difficult to generalize since privacy may still be implied in other information. Dave \textit{et al.} \cite{Dave_2022_CVPR} proposed a self-supervised learning to remove privacy information from videos without privacy labels, and meanwhile, it also supports action recognition. A general overview of the two schemes is shown in Fig. \ref{action recognition}, where it can be seen that both can achieve privacy-preserving action recognition, but the scheme \cite{Dave_2022_CVPR} does not require privacy labels.\par

Wu \textit{et al.} \cite{10.1145/3447993.3448618} designed a reversible transformation for video privacy protection supporting analytics, in which a combination of CycleGAN \cite{Zhu_2017_ICCV} and the key is utilized to ensure secure reversion. The experiment demonstrated that the scheme supports real-time operation, which enables the video recorded by the camera to be processed and transmitted to the analysis end in real time.\par

Generalized protection is nearly as powerful for privacy protection compared to target protection, erasing almost all visual information except that related to usability. On the other hand, in contrast to specific protection, this part of the scheme mostly requires adversarial training to enable CV models to perform specific tasks using the preserved information.\par

\begin{table*}[htb]
	\centering
	\caption{A summary of the solutions in general usability.}
	\renewcommand{\arraystretch}{1.5}
	\tabcolsep 7pt
	\label{tab: hv-GU}
	\begin{tabular}{ccccccccc}
		\toprule[2pt]
		\multirow{2}{*}{\makecell{Sub\\ class}} & \multirow{2}{*}{Paper} & \multirow{2}{*}{Year} & \multirow{2}{*}{\makecell{Key \\technology}} & \multirow{2}{*}{\makecell{Adversarial\\ training for task}} & \multirow{2}{*}{Reversibility} & \multicolumn{2}{l}{Type} & \multirow{2}{*}{CV task} \\ \cline{7-8}
		&                        &                       &                      &                  &              & I           & V          &                       \\ \hline
		
		\multirow{4}{*}{\makecell{Target\\protection}}  & \cite{9747456} & 2022 & \makecell{Postural\\ replacement} & \usym{2612} & \usym{2612} & \usym{2612} & \usym{2611} & \makecell{Person\\ segmentation} \\
		& \cite{Kunchala_2023_WACV} & 2023 & \makecell{3D\\ replacement} & \usym{2612} & \usym{2612} & \usym{2612} & \usym{2611} & \makecell{Pedestrian\\ analysis} \\
		& \cite{yang2022digital} & 2022 & \makecell{Digital\\ mask} & \usym{2612} & \usym{2612} & \usym{2611} & \usym{2612} & \makecell{Disease\\ diagnosis} \\
		& \cite{BRKIC201741} & 2017 & \makecell{Gaussian\\ mixture modes} & \usym{2612} & \usym{2612} & \usym{2612} & \usym{2611} & \makecell{Person\\ segmentation} \\
		\hline
		\multirow{5}{*}{\makecell{Generalized\\protection}}          &    \cite{Ryoo_Rothrock_Fleming_Yang_2017} &  2017 & \makecell{Inverse\\ super resolution} & \usym{2611} & \usym{2612} & \usym{2612} & \usym{2611} & \makecell{Action\\ recognition}\\
		&    \cite{Hinojosa_2021_ICCV} &  2021 & \makecell{Optical\\ encoder} & \usym{2611} & \usym{2612} & \usym{2611} & \usym{2612} & \makecell{Human pose\\ estimation}\\
		& \cite{9207852} & 2022 & \makecell{Supervised\\ learning} & \usym{2611} &  \usym{2612} & \usym{2612} & \usym{2611} & \makecell{Action\\ recognition} \\

		& \cite{Dave_2022_CVPR} & 2022 & \makecell{Self-supervised\\ learning} &\usym{2611} & \usym{2612} & \usym{2612} & \usym{2611} & \makecell{Action\\ recognition} \\
		& \cite{10.1145/3447993.3448618} & 2021 & CycleGAN & \usym{2612} & \usym{2611} & \usym{2612} & \usym{2611} & Detection \\
		\bottomrule[2pt]
	\end{tabular}
\end{table*}

\begin{table*}[htb]
	\centering
	\caption{A summary of the solutions in specific CV usability.}
	\renewcommand{\arraystretch}{1.5}
	\tabcolsep 7pt
	\label{tab: hv-su}
	\begin{tabular}{ccccccccc}
		\toprule[2pt]
		\multirow{2}{*}{\makecell{Sub\\ class}} & \multirow{2}{*}{Paper} & \multirow{2}{*}{Year} & \multirow{2}{*}{\makecell{Key \\technology}} & \multirow{2}{*}{\makecell{Adversarial \\ training for task}} & \multirow{2}{*}{Reversibility} & \multicolumn{2}{l}{Type} & \multirow{2}{*}{CV task} \\ \cline{7-8}
		&                        &                       &                      &                  &              & I           & V          &                       \\ \hline
		
		\multirow{5}{*}{\makecell{Person-\\related\\ usability}}  & \cite{Wu_2018_ECCV} & 2018 & Degradation & \usym{2611} & \usym{2612} & \usym{2612} & \usym{2611} & \makecell{Action\\ recognition} \\
		& \cite{Wang_2019_CVPR_Workshops} & 2019 & \makecell{Coded\\ aperture} & \usym{2611} & \usym{2612} & \usym{2612} & \usym{2611} & \makecell{Action\\ recognition} \\
		& \cite{9220826} & 2021 & \makecell{Compressed\\ sensing} & \usym{2611} & \usym{2612} & \usym{2612} & \usym{2611} & \makecell{Fall\\ detection} \\
		& \cite{9666933} & 2021 & Masking & \usym{2611} & \usym{2612} & \usym{2611} & \usym{2612} & \makecell{Face\\ recognition} \\
		& \cite{10.1007/978-3-031-19775-8_28} & 2022 & \makecell{Frequency\\ domain transform} & \usym{2611} & \usym{2612} & \usym{2611} & \usym{2612} & \makecell{Face\\ recognition} \\
		\hline
		\multirow{3}{*}{\makecell{Other}}          &    \cite{10.1145/3442381.3449907} &  2021 & \makecell{Reducing\\ pixel entropy} & \usym{2611} & \usym{2612} & \usym{2611} & \usym{2612} & Inference\\
		&    \cite{10.1145/3524273.3528189} &  2022 & Perturbation & \usym{2612} & \usym{2612} & \usym{2611} & \usym{2612} & Recognition\\
		& \cite{9366496} & 2021 & Encryption & \usym{2611} &  \usym{2612} & \usym{2611} & \usym{2612} & Classification \\
		\bottomrule[2pt]
	\end{tabular}
\end{table*}

\subsection{Solutions for Specific CV Usability}

\subsubsection{Person-related usability}\

Wu \textit{et al.} \cite{Wu_2018_ECCV} proposed an adversarial training framework to achieve privacy-preserving CV recognition, in which the original video is transformed into a visually almost imperceptibly degraded version. The framework explicitly learns the degraded transform for the original video inputs to optimize the trade-off between the target performance and the privacy budget (introducing differential privacy) on the degraded video. Experiments showed that the risk of privacy leakage is effectively suppressed while action recognition is achieved. Wang \textit{et al.} \cite{Wang_2019_CVPR_Workshops} proposed a lens-free coded aperture camera system for privacy protection while enabling the human action recognition, where the protected visual content is unintelligible and does not reveal any information. This scheme applies the coded aperture principle \cite{cannon1980coded}, which is capable of erasing visual features, while the authors utilized deep neural networks to use the coded aperture data for action recognition. Liu \textit{et al.} \cite{9220826} proposed a fall detection system with visual shielding, which can effectively detect the occurrence of a fall while achieving a home camera that cannot record additional information. It implements visual shielding by multi-layer compressed sensing, then proposes a corresponding feature extraction algorithm to extract human behavior features, and finally implements fall detection based on GAN.\par

Facial regions have been the focus of research on privacy-preserving and usability recognition, and there is also some work supporting specific CV usability that focuses on facial regions. Gupta \textit{et al.} \cite{9666933} proposed that when performing a specific CV task, only locally necessary information relevant to the task is required, while all other information can be erased. Experiments showed that it can successfully perform facial emotion and attribute recognition after erasing all non-essential information. Ji \textit{et al.} \cite{10.1007/978-3-031-19775-8_28} transformed the original image to the frequency domain and then removed the DC coefficients to add random perturbation in the framework of differential privacy to provide strong protection for facial privacy. Meanwhile, adversarial training is performed to achieve face recognition in the privacy-preserving case.\par

Person-related usability has been the focus of CV research. The above studies have tried their best to remove visual information while preserving person-related CV usability. However, it is important to note that while person-related usability/privacy is a hot topic, there is a lot of other visual content that is also integral to CV usability. 

\subsubsection{Other}\

Outsourcing visual content data for AI-related detection and inference is one of the hot directions, but it also raises the risk of data privacy, e.g., data may be used for unintended tasks. Wu \textit{et al.} \cite{10.1145/3442381.3449907} proposed to minimize the unnecessary pixel entropy before delivering the image data for outsourcing. They used a neural network to train a model, and the trained model can transform the input image to remove irrelevant information on demand, while ensuring good inference accuracy for a specific CV task. Ye \textit{et al.} \cite{10.1145/3524273.3528189} proposed a protective perturbation generator to protect the privacy of outsourced images, which can effectively eliminate the visual information in the images and prevent privacy leakage without affecting the target recognition model. Meanwhile, it does not require the outsourced model to be retrained. In addition, Aprilpyone \textit{et al.} \cite{9366496} proposed three image transformation algorithms based on cryptographic ideas.  These protected images are utilized for adversarial training and the resulting model is able to classify the images effectively. Meanwhile, the visual information contained in the image is effectively protected or even erased.\par

Overall, this part mitigates the privacy risks associated with outsourcing visual content data while preserving the usability of specific tasks (e.g., recognition and classification).
This may make people more willing to share visual content with third-party services, and to some extent may help solve the data silo problem. However, they are often targeted at coarse-grained CV tasks and may be powerless for more detailed tasks, such as face recognition.\par

\subsection{Common Principles}

Tables \ref{tab: hv-GU} and \ref{tab: hv-su} provide the breakdown of the reviewed solutions in general usability protection and specific CV usability, respectively, in which adversarial training for tasks means that the model needs to be trained on protected visual content in order to perform a specific CV task.\par

By reviewing and summarizing the above solutions, 3 common principles can be identified.\par

\textbf{Abstraction.} Since this part considers not only the defense against HV and CV adversaries, but also the usability of CV, it is often necessary to extract the valid information from the visual content while other unnecessary information is removed. It can be seen as a high level of abstraction of visual content.\par

\textbf{Confidential level.} As mentioned above, this part of the protection of the visual content tends to remove all unnecessary information, leaving only the key information necessary to perform the CV task. Therefore, for almost protection schemes, it is difficult to obtain much useful information about privacy content from HV alone, let alone for CV tasks and models that are not expected.\par

\textbf{Customization.} CV usability in this part is often a customization, that is, a customized protection of the visual content in order to take a specific task, ensuring that the information left is sufficient to perform that task. Even the CV models used need to be trained adversarially based on the protected visual content to obtain a customized model for performing that task.\par

\section{Challenges and Future Directions} \label{Section: challenges}

There is no doubt that privacy has been taken very seriously across all communities, from legal \cite{voigt2017eu} to engineering \cite{4657365}, from politics \cite{mokrosinska2018privacy} to science \cite{anderlik2001privacy}, and a number of visual privacy solutions are rapidly being proposed to address or mitigate various privacy risks. However, there are several foundational or important problems about privacy protection that need to be considered as priorities. In this section, we summarize what we see as the main challenges related to the privacy protection of visual content.\par

\subsection{What is privacy?}

All privacy protection schemes talk about privacy, but they tend to assume that something is private, such as faces, and that processing it is the same as completing privacy protection. However, is this really a privacy protection? Or is it just an objective visual content processing task? These schemes do not care about defining the concept of privacy, but go straight to talking about privacy protection, and thus the protection of privacy is often difficult to put in place.\par

In fact, there is still no universal conclusion about the concept of privacy. Just as Thomson said \cite{10.2307/2265075} that ``\textit{Perhaps the most striking thing about the right to privacy is that nobody seems to have any clear idea of what it is}''. Privacy is significantly cross-disciplinary and interdisciplinary \cite{6890932}. Research on privacy involves sociology, law, political science, philosophy, ethics, economics, etc \cite{7436680}. These studies and discourses differ significantly in their definitions of the concept of privacy, making the findings vary widely and even fundamentally opposed.\par

While some scholars have attempted to provide some definitions of privacy in different dimensions, it tends to be highly generalized. For example, a scholar \cite{10.1145/293411.293475} argued that ``\textit{Interest that individuals have in sustaining personal space, free from interference by other people and organizations}''. This may seem right, but it is difficult to use it to clarify what is privacy in practice. \par

If we can clarify what privacy is, it can bring a huge improvement to privacy protection in at least 3 ways. \textit{1)} Privacy can be fully, automatically, and adaptively protected, rather than sporadically for a particular attribute or object, as it is today. \textit{2)} Privacy protection can be measured, quantified, and even made into a science, rather than an art that can only be verified using extensive experiments. \textit{3)} A theoretically optimal balance between privacy protection and usability can be further approached without worrying that too much protection will spill over into irrelevant usability and too little protection will not eliminate privacy.\par

\subsection{Privacy and Usability}

Some scholars pointed out that privacy should be a personal interest \cite{10.2307/23415061,10.1145/293411.293475}. This means that privacy needs to be a trade-off against other interests, which may be personal, political, social, or technological \cite{9481149}. In fact, privacy protection is not often the primary goal (which is usability) of the user, but a secondary one \cite{cranor2005security}. Privacy protection for visual content is meaningless if it completely compromises usability. For example, the purpose of a user posting an image on an OSN is for a certain visual viewability for others. If this viewability is completely erased, it ensures that the visual content does not compromise privacy, but then does the user still need to post the image? \par

The trade-off between data privacy and usability is always an open challenge for privacy protection \cite{10.1145/3547299}, especially for visual content because of the diversity and complexity of information it contains. In general, the greater the privacy, the worse the usability, and vice versa, as shown in Fig. \ref{TPE-diff}. Ideally, privacy protection and usability should be separated from each other. That is, only private information is processed, without affecting any extraneous information to provide usability.\par

There are two difficulties in this separation of privacy and usability. \textit{1)} In theory, what privacy is is still inconclusive as mentioned above, and it is difficult to separate things that cannot be defined. \textit{2)} In technology, it is difficult to extract information accurately as well as to process the target information without affecting others.\par

It should be noted that researchers have been advancing in terms of technology. Especially on the face, precise and controlled extraction, processing, and synthesis techniques have been investigated \cite{Deng_2020_CVPR,Ren_2021_ICCV,10.1145/3550454.3555506}. However, they still inevitably change some attributes irrelevant to the target (privacy), but of course, they are more precise than previous work.\par

\subsection{Recognition of Protection Area}

Visual content privacy protection either modifies all the areas or circles only one (or more) ROI(s) to modify. Almost all ROI protection reviewed in this survey either assumes that this is related to a particular object, such as face, or is directly hand-drawn. In fact, the privacy objects in different visual contents may not be the same and are very diverse. Meanwhile, humans' perception of privacy in visual content is often not that comprehensive, thus leading to making privacy decisions about ROI that are often prone to errors. It is also very difficult and time-consuming for humans to manually label privacy for all visual content \cite{10.1145/3386082}.\par

Some privacy prediction methods have also been proposed to automate the process of helping humans to identify the privacy of visual content \cite{10.1145/3386082,Orekondy_2017_ICCV} or to give privacy labels \cite{10.1145/3335054}. However, these methods often still have a huge database marked with what is private content and what is otherwise \cite{10.1145/3335054}. Alternatively, privacy is assessed based on pre-specified privacy attributes \cite{,Orekondy_2017_ICCV,10.1145/3386082}. This means that if it is not in databases, or if a certain type of information is not pre-specified, then it cannot be correctly predicted either.\par

It is also important to note that the existing non-manual privacy recognition may have a certain error rate. If privacy is not properly recognized as well as circled, then privacy protection will fail. Meanwhile, for video, if one frame is not circled, then the rest of the effect is greatly diminished. Therefore, how to reduce the probability of CV misidentifying privacy areas and improve the robustness of detection, while recognizing as many kinds of privacy as possible is crucial to automate the recognition of protected areas.\par

\subsection{People-Oriented Protection}

The topic of privacy has always been of great interest, and images and videos in particular, because of the rich and visual information they can carry, have led to a constant stream of advanced schemes for visual privacy protection. On the other hand, in real life, it may be rare to see so many advanced schemes in use, and the vast majority of ordinary people (non-academics) are not even aware of them. They still utilize seemingly backward ways (pixelation and blurring) to protect visual privacy. Even, many of the more recently proposed advanced schemes (e.g., \cite{WEN2022197,10.1145/3503161.3547757,10.1145/3503161.3548235}) still choose these age-old methods as a comparison since they are simply too common.\par

Compared to advanced schemes, pixelation/blurring has two huge advantages: 1) no need for the user to master any expertise, anyone can simply use it to get the obvious protection effect; 2) no high computational cost, ordinary devices can be executed in real time \cite{rajabi2021practicality}. These two points can be described as a kind of people-oriented protection. Many advanced schemes do not consider the people-oriented characteristics, especially for CV adversaries. For example, adversarial perturbation schemes can be divided into white-box and black-box. The white-box schemes require the user to have the full knowledge of the adversarial CV model, including structure, parameters, and output classes. This is not to mention the ordinary people, professionals may also be difficult to fully know the knowledge of the adversary. The black-box schemes require extensive testing, which entails a large computational cost that is difficult to support with the ordinary devices that the user has, and they also require the user to have expertise in training the model.\par

If this people-oriented characteristic is implemented in a protection scheme, it may help the scheme to become rapidly popular in daily life. Meanwhile, it should be noted that privacy protection itself should be people-oriented, since it is intended to help the individual to protect privacy. If the scheme is not easily used by individuals but requires expertise and a lot of effort, it is undesirable and even difficult to accept in practice, even if it has excellent results in the laboratory.\par

\section{Conclusion} \label{Section: conclusion} 

Vision is an important source of human perception of the world, always faster and more intuitive than other ways of obtaining information. Therefore, visual content always conveys information quickly and accurately, and thus people are willing to use visual content data to record and share their lives. However, the convenience of visual content comes with serious privacy concerns and is only expected to increase over time. A great deal of research has been done to address or mitigate these concerns, making the benefits of visual content usable to individuals while also taking steps to minimize the privacy impact of visual content. That is, these researches attempt to strike a balance between privacy and usability of visual content.\par

This survey aims to provide a comprehensive overview of research related to the privacy protection of visual content. Visual content privacy protection is the collective term for solutions that address or mitigate the privacy concerns that arise from the visual effects of image and video content. First, the characteristics of visual content privacy and the difficulty and challenge of protection are introduced. Then, for the criterion that privacy protections have adversaries, a high-level privacy protection framework based on the type of adversary is proposed. Any privacy protection scheme will be able to find a corresponding classification in this framework. For each of these classifications, we reviewed the corresponding solutions, summarized and compared them, and also proposed their common principles. Finally, the open challenges and future directions for visual privacy protection are discussed.\par

This survey explores a comprehensive picture of visual privacy protection, and privacy adversaries and solutions are included in this survey. We hope that our work will provide a macro perspective on current research, contribute to the development of the visual privacy protection community, and provide assistance for individual privacy protection needs.\par

Finally, we also suggest to service providers (e.g., cloud, surveillance, and CV's API) who process visual content that they take privacy protection seriously. For the public, their awareness of privacy continues to grow, and they are increasingly intolerant of a service with privacy concerns. For example, in 2021, Apple tried to promote the detection of images of child sexual violence on iCloud \cite{gernand2022scanning}, which sounded reasonable. However, many people immediately attacked it since it was, in fact, a kind of surveillance system on the device, which seriously violated the privacy of users. For this reason, the system was announced to be canceled by Apple before it had a chance to go live. This means that if service providers do not take privacy seriously, their products may be boycotted by users before they even have a chance to get distributed. Meanwhile, laws, regulations, and standards, e.g., EU's GDPR \cite{voigt2017eu}, Illinois's Biometric Information Privacy Act \cite{insler2018ride}, and ISO/IEC 30137-1:2019 \cite{ISO/IEC30137-1}, are also becoming clearer and stricter on privacy issues, with less and less ambiguity. This also means that privacy violations can lead to investigations, questioning, and large fines in some countries. In short, privacy should now be rapidly gaining ground in service providers, and while it may not make a mediocre product successful, it can make a good product discarded by both the public and the government.

\bibliographystyle{IEEEtran}
\bibliography{ref}

\end{document}